\documentclass[letterpaper, 10pt, conference]{ieeeconf}      

\IEEEoverridecommandlockouts                              
  \pdfoutput=1                                                        

\usepackage{graphics} 
\usepackage{epsfig} 
\usepackage{times} 
\usepackage{amsmath} 
\usepackage{amssymb}  
\usepackage{subfigure}
\usepackage{url}
\usepackage{bbm, dsfont}
\usepackage{mathtools}
\usepackage{multirow}
\newtheorem{theorem}{Theorem}

\newtheorem{algorithm}[theorem]{Algorithm}

\usepackage{cite}

\usepackage{color}
\usepackage{algorithm}
\newtheorem{properties}[theorem]{Properties}

\newtheorem{definition}[theorem]{Definition}

{ 
\newtheorem{remark}[theorem]{Remark}

}

\newcommand{\bi}{\begin{itemize}}
\newcommand{\ei}{\end{itemize}}
\newcommand{\bd}{\begin{displaymath}}
\newcommand{\ed}{\end{displaymath}}
\newcommand{\be}{\begin{eqnarray*}}
\newcommand{\ee}{\end{eqnarray*}}

\newcommand{\K}{{\bf K}}
\usepackage[normalem]{ulem}

\title{\LARGE \bf
Online Learning of Dynamical Systems: An Operator Theoretic Approach}
\author{Subhrajit Sinha, Sai Pushpak Nandanoori and Enoch Yeung\\
\thanks{The Pacific Northwest National Laboratory (PNNL) is operated by Battelle for the U.S. Department of Energy under Contract DE-AC05-76RL01830.}
\thanks{S. Sinha and S. P. Nandanoori are with PNNL, Richland, WA 99354 USA (emails: subhrajit.sinha@pnnl.gov, saipushpak.n@pnnl.gov,  and E. Yeung is with University of California Santa Barbara, CA 93106 (email: eyeung@ucsb.edu)
}
}

\begin{document}
\maketitle

\begin{abstract}
In this paper, we provide an algorithm for online computation of Koopman operator in real-time using streaming data. In recent years, there has been an increased interest in data-driven analysis of dynamical systems, with operator theoretic techniques being the most popular. Existing algorithms, like Dynamic Mode Decomposition (DMD) and Extended Dynamic Mode Decomposition (EDMD), use the entire data set for computation of the Koopman operator. However, many real life applications like power system analysis, biological systems, building systems etc. requires the real-time computation and updating of the Koopman operator, as new data streams in. In this paper, we propose an iterative algorithm for online computation of Koopman operator such that at each time step the Koopman operator is updated incrementally. In particular, we propose a Recursive Extended Dynamic Decomposition (rEDMD) algorithm for computation of Koopman operator from streaming data. Further, we test the algorithm in three different dynamical systems, namely, a linear system, a nonlinear system and a system governed by a Partial Differential Equation (PDE) and illustrate the computational efficiency of the iterative algorithm over the existing DMD and EDMD algorithms.
\end{abstract}

\section{Introduction}\label{section_introduction}

Analysis and control of dynamical systems is a matured branch of science and engineering with applications in various branches of science and engineering. In the past, much of the research in applications of dynamical systems theory had been model based, where an \emph{a priori} knowledge of the mathematical model of the system concerned had to be assumed. But in many applications like power networks and biological systems, to mention a few, it is extremely difficult to obtain a good enough model for the underlying dynamics. However, in such applications, it is often easier to acquire data from the system, often in the form of values of the states of the system. This requires a framework for data-driven analysis of dynamical systems. 

In recent years, there has been an increased interest in transfer operator based analysis and control of dynamical systems \cite{Dellnitz_Junge,Mezic2000,froyland_extracting,Junge_Osinga,Mezic_comparison,
Dellnitztransport,mezic2005spectral,Mehta_comparsion_cdc,Vaidya_TAC,
raghunathan2014optimal,susuki2011nonlinear,mezic_koopmanism,
mezic_koopman_stability,surana_observer,yeung2015global, yeung2018koopman,yeung2017learning, sparse_Koopman_acc,johnson2018class}. The operator based approach is fundamentally different from the classical approach in the sense that instead of studying the dynamical system on the configuration manifold and its tangent and cotangent bundle, the system is lifted to the function space or to the space of measures and is studied there. Though the operators, that govern the evolution of the system in the function space or measure space, are usually infinite dimensional, the advantage lies in the fact that in those spaces, the evolution is linear, even if the actual system is nonlinear. Moreover, the operators, namely, Perron-Frobenius (P-F) operator and the Koopman operator are positive Markov operators which can be exploited to have probabilistic interpretations and can be used for various applications like the optimal placement of sensors and actuators \cite{optimal_placement_ECC, optimal_placement_JMAA} etc.

Apart from the system evolution being linear in the lifted space, another major advantage of the operator theoretic approach is the fact that it facilitates the data-driven analysis of dynamical systems. In particular, approximations of both P-F and Koopman operators can be efficiently computed from time-series data, obtained from simulation or from experiments. To this end, various methods have been proposed to compute the approximations of these operators from data \cite{dellnitz2002set, Mezic2000,DMD_schmitt,rowley2009spectral,EDMD_williams}, with Dynamic Mode Decomposition (DMD) and Extended Dynamic Decomposition (EDMD) being the most popular ones.  Recent researches have also addressed the problem of computing these operators for systems with process and observation noise and for Random Dynamical Systems (RDS)  \cite{mezic_stochastic_koopman_spectrum,PhysRevE.96.033310, robust_DMD_ACC,robust_DMD_arxiv}.

In all the above approaches, the approximate Koopman operator is computed using the entire obtained data-set, that is, using batch data-sets. However, in many different applications, like real-time analysis and control of power networks or building systems, biological systems etc., it is necessary to construct the approximate dynamics in real-time. This requires a framework which will facilitate the computation of Koopman operator from streaming data. Further, in many cases these systems may be of extremely large dimensions, and so recomputing the Koopman operator every time a each new data point streams in, will be computationally expensive. However, these issues can be addressed if an iterative algorithm can be used for Koopman operator computation, such that the Koopman operator computed at the previous time step can be incrementally updated to give the new Koopman operator. 

In this paper, we propose an algorithm that computes the Koopman operator recursively, such that the algorithm at each time step takes one time step data at a time as input and updates the Koopman operator which was obtained from the previous time step. In particular, we provide a recursive version of the popular EDMD algorithm such that online computation of Koopman operator can be achieved. 

The paper is organized as follows. In section \ref{section_premilinaries} we discuss the basic theory of transfer operators followed by a brief introduction to DMD and EDMD algorithms in section \ref{section_DMD}. In section \ref{section_online} we derive the recursive Koopman computation algorithm. Design of predictor using the online Koopman operator is presented in section \ref{section_predictor} followed by simulation results in section \ref{section_simulation}. Finally the paper is concluded in section \ref{section_conclusion}.

\section{Preliminaries}\label{section_premilinaries}
Consider a discrete-time dynamical system
\begin{eqnarray}\label{system}
z_{t+1} = T(z_t)
\end{eqnarray}
where $T:Z\subset \mathbb{R}^N \to Z$ is assumed to be an invertible smooth diffeomorphism.  Associated with the dynamical system (\ref{system}) is the Borel-$\sigma$ algebra ${\cal B}(Z)$ on $Z$ and the vector space ${\cal M}(Z)$ of bounded complex valued measures on $X$. With this, two linear operators, namely, Perron-Frobenius (P-F) and Koopman operator, can be defined as follows \cite{Lasota} :
\begin{definition}[Perron-Frobenius Operator] 
The P-F operator $\mathbb{P}:{\cal M}(Z)\to {\cal M}(Z)$ is given by

{\small
\begin{eqnarray}
[\mathbb{P}\mu](A)=\int_{{\cal Z} }\delta_{T(z)}(A)d\mu(z)=\mu(T^{-1}(A))
\end{eqnarray}
}
where $\delta_{T(z)}(A)$ is stochastic transition function which measure the probability that point $z$ will reach the set $A$ in one time step under the system mapping $T$. 
\end{definition}



\begin{definition} [Koopman Operator] 
Given any $h\in\cal{F}$, $\mathbb{U}:{\cal F}\to {\cal F}$ is defined by
\[[\mathbb{U} h](z)=h(T(z))\]
where $\cal F$ is the space of function (observables) invariant under the action of the Koopman operator.
\end{definition}

\begin{figure}[htp!]
\centering
\includegraphics[scale=.25]{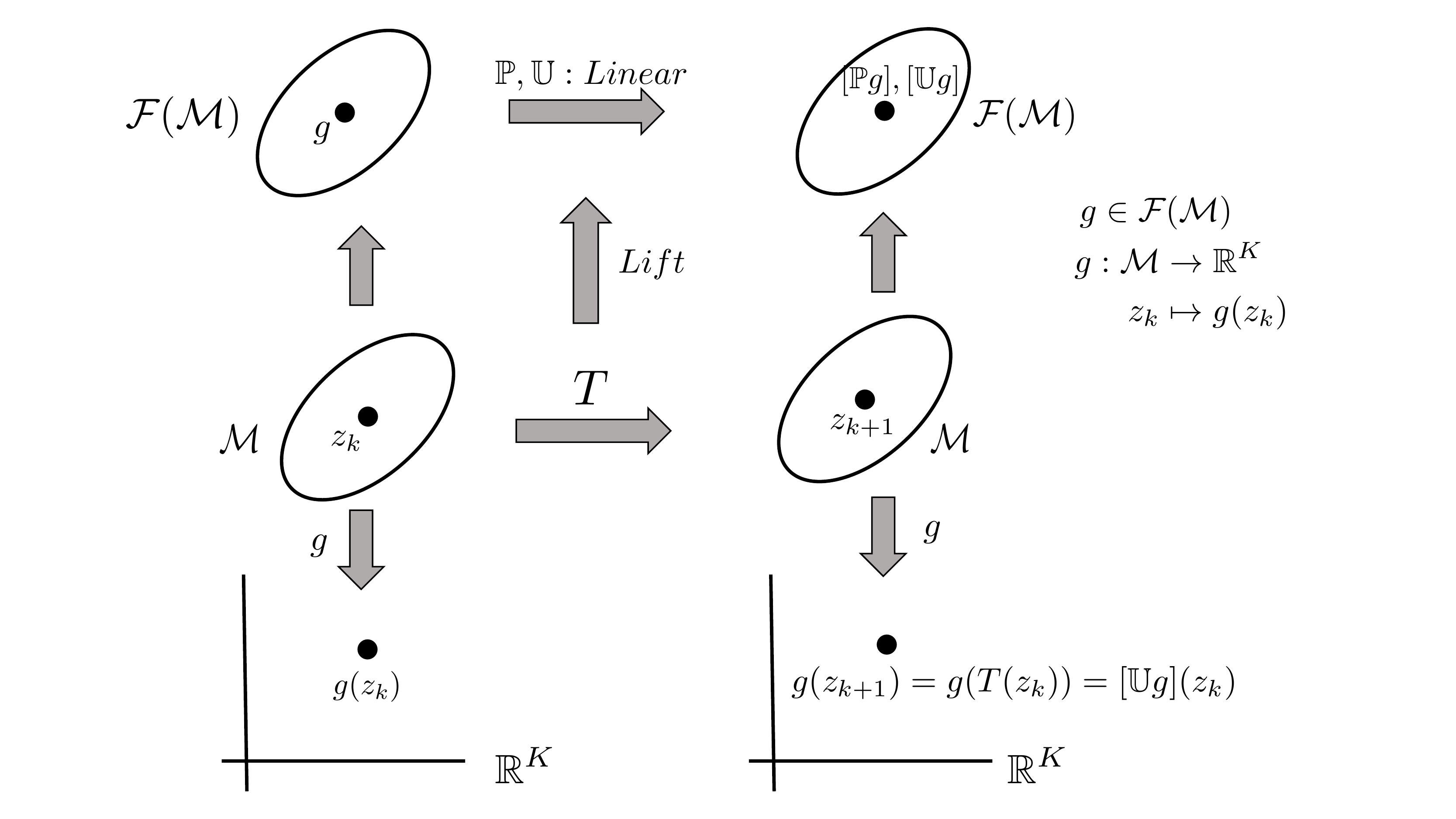}
\caption{Schematic of the P-F and Koopman operators.}\label{koopman_diagram}
\end{figure}

Both the P-F operator and the Koopman operators are linear operators, even if the underlying system is non-linear. But while analysis is made tractable by linearity, the trade-off is that these operators are typically infinite dimensional. In particular, the P-F operator and Koopman operator often will lift a dynamical system from a finite-dimensional space to generate an infinite dimensional linear system in infinite dimensions (see Fig. \ref{koopman_diagram}).

\begin{properties}\label{property}
The following properties for the Koopman and Perron-Frobenius operators can be stated \cite{Lasota}:

\begin{enumerate}
\item [a).] For the Hilbert space ${\cal F}=L_2(Z,{\cal B}, \bar \mu)$, on measure preserving systems, the Koopman operator $\mathbb{U}$ satisfies
\begin{eqnarray*}
&&\parallel \mathbb{U}h\parallel^2=\int_Z |h(T(z))|^2d\bar \mu(z)
\nonumber\\&=&\int_Z | h(z)|^2 d\bar\mu(z)=\parallel h\parallel^2
\end{eqnarray*}
where $\bar \mu$ is an invariant measure. This implies that Koopman operator is unitary for measure preserving systems.

\item [b).] For any $h\geq 0$, $[\mathbb{U}h](z)\geq 0$ and hence the Koopman operator is a positive operator.

\item [c).]For invertible system $T$, the P-F operator for the inverse system $T^{-1}:Z\to Z$ is given by $\mathbb{P}^*$ and $\mathbb{P}^*\mathbb{P}=\mathbb{P}\mathbb{P}^*=I$. Hence, the P-F operator is unitary.

\item [d).] If the P-F operator is defined to act on the space of densities i.e., $L_1(Z)$ and Koopman operator on space of $L_\infty(Z)$ functions, then it can be shown that the P-F and Koopman operators are dual to each other \footnote{with some abuse of notation we use the same notation for the P-F operator defined on the space of measure and densities.}
\begin{eqnarray*}
&&\left<\mathbb{U} f,g\right>=\int_Z [\mathbb{U} f](z)g(z)dz\nonumber\\&=&\int_Z f(y)g(T^{-1}(y))\left|\frac{dT^{-1}}{dy}\right|dy=\left<f,\mathbb{P} g\right>
\end{eqnarray*}
where $z=Ty$, $f\in L_{\infty}(Z)$ and $g\in L_1(Z)$ and the P-F operator on the space of densities $L_1(Z)$ is defined as follows
\[[\mathbb{P}g](z)=g(T^{-1}(z))\left|\frac{dT^{-1}(z)}{dz}\right|.\]

\item [e).] For $g(z)\geq 0$, $[\mathbb{P}g](z)\geq 0$.

\item [f).] Let $(Z,{\cal B},\mu)$ be the measure space where $\mu$ is a positive but not necessarily the invariant measure of $T:Z\to Z$, then the P-F operator $\mathbb{P}:L_1(Z,{\cal B},\mu)\to L_1(Z,{\cal B},\mu)$  satisfies  following property:

 \[\int_Z [\mathbb{P}g](z)d\mu(z)=\int_Z g(z)d\mu(z).\]\label{Markov_property}
\end{enumerate}
\end{properties}

\section{Dynamic Mode Decomposition (DMD) and Extended Dynamic Mode Decomposition (EDMD)}\label{section_DMD}

Dynamic Mode Decomposition, initially developed in \cite{DMD_schmitt} in the context of analysis of fluid flow analysis, is a method for approximating the spectrum of the Koopman operator. In \cite{EDMD_williams}, the authors generalized the DMD algorithm to what is now known in literature as Extended Dynamic Mode Decomposition (EDMD). In EDMD algorithm, the Koopman operator is approximated as a linear map on the span of a finite set of dictionary functions. In this section, we briefly describe the EDMD algorithm for approximating the Koopman operator. 

Consider snapshots of data set 
\begin{eqnarray}
X_p = [x_1,x_2,\ldots,x_M],& X_f = [y_1,y_2,\ldots,y_M] \label{data}
\end{eqnarray}
obtained from simulating a discrete time dynamical system $x\mapsto T(x)$, $x\in X\subset \mathbb{R}^N$, or from an experiment, where $x_i\in X$ and $y_i\in X$. The two pair of data sets are assumed to be two consecutive snapshots i.e., $y_i=T(x_i)$. Let $\mathcal{D}=
\{\psi_1,\psi_2,\ldots,\psi_K\}$ be the set of dictionary functions or observables, where $\psi : X \to \mathbb{C}$. Let ${\cal G}_{\cal D}$ denote the span of ${\cal D}$ such that ${\cal G}_{\cal D}\subset {\cal G}$, where ${\cal G} = L_2(X,{\cal B},\mu)$. Define vector valued function $\mathbf{\Psi}:X\to \mathbb{R}^{K}$
\begin{equation}
\mathbf{\Psi}(\boldsymbol{x}):=\begin{bmatrix}\psi_1(x) & \psi_2(x) & \cdots & \psi_K(x)\end{bmatrix}^\top.
\end{equation}
In this application, $\mathbf{\Psi}$ is the mapping from physical space to feature space. Any function $\phi,\hat{\phi}\in \mathcal{G}_{\cal D}$ can be written as
\begin{eqnarray}
\phi = \sum_{k=1}^K a_k\psi_k=\boldsymbol{\Psi^T a},\quad \hat{\phi} = \sum_{k=1}^K \hat{a}_k\psi_k=\boldsymbol{\Psi^T \hat{a}}
\end{eqnarray}
for some set of coefficients $\boldsymbol{a},\boldsymbol{\hat{a}}\in \mathbb{R}^K$. Let \[ \hat{\phi}(x)=[\mathbb{U}\phi](x)+r,\]
where $r$ is a residual function that appears because $\mathcal{G}_{\cal D}$ is not necessarily invariant to the action of the Koopman operator. The finite dimensional approximate Koopman operator $\bf K$ minimizes this residual $r$ and the matrix $\bf K$ is obtained as a solution of the following least square problem: 
\begin{equation}\label{edmd_op}
\min\limits_{\bf K}\parallel{\bf K} {Y_p}-{Y_f}\parallel_F
\end{equation}
where
\begin{eqnarray}\label{edmd1}
\begin{aligned}
& {Y_p}={\bf \Psi}(X_p) = [{\bf \Psi}(x_1), {\bf \Psi}(x_2), \cdots , {\bf \Psi}(x_M)]\\
& {Y_f}={\bf \Psi}(X_f) = [{\bf \Psi}(y_1), {\bf \Psi}(y_2), \cdots , {\bf \Psi}(y_M)],
\end{aligned}
\end{eqnarray}
with ${\bf K}\in\mathbb{R}^{K\times K}$. The optimization problem (\ref{edmd_op}) can be solved explicitly to obtain following solution for the matrix $\bf K$
\begin{eqnarray}
{\bf K}={Y_f}{Y_p}^\dagger \label{EDMD_formula}
\end{eqnarray}
where ${Y_p}^{\dagger}$ is the pseudo-inverse of matrix $Y_p$.
DMD is a special case of EDMD algorithm with ${\bf \Psi}(x) = x$.

\section{Online Computation of Koopman Operator}\label{section_online}

The Koopman operator is generally computed by solving the optimization problem (\ref{edmd_op}) or directly from the formula (\ref{EDMD_formula}), where one uses the entire dataset for the computation. Hence, if some new data point is acquired, a new Koopman operator is computed using the new enlarged data-set. However, this requires inversion of a matrix and this is computationally expensive, specially for data-sets obtained from a large dimensional system with a huge number of data points. This warrants a recursive algorithm for Koopman operator computation.

In this section, we describe an algorithm which computes the Koopman operator recursively and thus reducing the computational cost. 
Let
\begin{eqnarray}
^MX_p = [x_1,x_2,\ldots,x_M],& ^MX_f = [y_1,y_2,\ldots,y_M] \label{data_m}
\end{eqnarray}
be $M$ data points obtained from simulation of a dynamical system $x\mapsto T(x)$ or from an experiment, where $y_i=T(x_i)$. Let
\begin{eqnarray}\label{data_m_lifted}
\begin{aligned}
& ^M{Y_p}={\bf \Psi}(X_p) = [{\bf \Psi}(x_1), {\bf \Psi}(x_2), \cdots , {\bf \Psi}(x_M)]\\
& ^M{Y_f}={\bf \Psi}(X_f) = [{\bf \Psi}(y_1), {\bf \Psi}(y_2), \cdots , {\bf \Psi}(y_M)],
\end{aligned}
\end{eqnarray}
be the data points in the lifted space $(\mathbb{R}^K)$, where the points $x_i$ and $y_i$ are mapped by the dictionary functions ${\bf \Psi}$. Let
\begin{eqnarray}\label{Koopman_m_step}
{\bf K}_M = ^M{Y_f}^M{Y_p}^\dagger
\end{eqnarray}
be the Koopman operator obtained by using the $M$ data points. Now, a new data point $(x_{M+1},y_{M+1})$ is aquired. The problem is to update the Koopman operator $\K_M$ to ${\bf K}_{M+1}$, without explicitly computing the inverse $(^{M+1}{Y_p})^\dagger$.

Note that (\ref{Koopman_m_step}) can be rewritten as
\begin{eqnarray}
\K_M \phi_M = z_M
\end{eqnarray}
where 
\begin{eqnarray}
\begin{aligned}
& \phi_M = ^MY_p(^MY_p)^\top=\sum_{i=1}^MY_p^i(Y_p^i)^\top\\
& z_M = ^MY_f(^MY_p)^\top=\sum_{i=1}^MY_f^i(Y_p^i)^\top
\end{aligned}
\end{eqnarray}
and $Y_p^i$ and $Y_f^i$ are $i^{th}$ columns of $^MY_p$ and $^MY_f$ respectively.
Moreover, the updated Koopman operator $\K_{m+1}$ satisfies
\begin{eqnarray}\label{Koopman_updated}
\K_{M+1}\phi_{M+1} = z_{M+1}
\end{eqnarray}
where
\begin{eqnarray}
\begin{aligned}
\phi_{M+1} = ^{M+1}Y_p(^{M+1}Y_p)^\top=\sum_{i=1}^{M+1}Y_p^i(Y_p^i)^\top\\
z_{M+1}=^{M+1}Y_f(^{M+1}Y_p)^\top=\sum_{i=1}^{M+1}Y_f^i(Y_p^i)^\top.
\end{aligned}
\end{eqnarray}

Now,
\begin{eqnarray*}
\phi_{M+1} &=& \sum_{i=1}^{M+1}Y_p^i(Y_p^i)^\top\\
&=& \sum_{i=1}^{M}Y_p^i(Y_p^i)^\top + Y_p^{M+1}(Y_p^{M+1})^\top\\
&=& \phi_M + Y_p^{M+1}(Y_p^{M+1})^\top.
\end{eqnarray*}
Hence, using the Matrix Inversion Lemma, we have
\begin{eqnarray}\label{phi_iterate}
\phi_{M+1}^{-1} = \phi_M^{-1} - \frac{\phi_M^{-1}Y_p^{M+1}(Y_p^{M+1})^\top\phi_M^{-1}}{1 + (Y_p^{M+1})^\top \phi_M^{-1} Y_p^{M+1}}.
\end{eqnarray}
Moreover,
\begin{eqnarray}\label{zm_iterate}
z_{M+1} = \sum_{i=1}^{M+1}Y_f^i(Y_p^i)^\top=z_M + Y_f^{M+1}(Y_p^{M+1})^\top.
\end{eqnarray}
Hence, from (\ref{Koopman_updated}),

\begin{eqnarray}\label{Koopman_new}\nonumber
\K_{M+1} &=& z_{M+1}\phi_{M+1}^{-1}\\ \nonumber
&=& \left(z_M + Y_f^{M+1}(Y_p^{M+1}\right)^\top)\times\\
&& \left(\phi_M^{-1} - \frac{\phi_M^{-1}Y_p^{M+1}(Y_p^{M+1})^\top\phi_M^{-1}}{1 + (Y_p^{M+1})^\top \phi_M^{-1} Y_p^{M+1}}\right).
\end{eqnarray}

Equation (\ref{Koopman_new}) gives the formula for updating the Koopman operator as new data streams in, without explicitly computing the inverse at every step, thus reducing the computational cost and hence improving efficiency.

\subsection{Initialization of the Algorithm}

Equation (\ref{Koopman_new}) gives the updates Koopman $\K_{M+1}$ operator in terms of quantities computed from the previous time step. Note that the updated Koopman operator $\K_{M+1}$ depends on an inverse, namely, $\phi_M^{-1}$. Hence, for computing the Koopman operator $\K_1$, one needs to initialize both $\phi_0$ and $z_0$. One potential way out of this situation is to compute the Koopman operator $\K_q$ using the initial $q$ data points $(x_i,y_i)$, $i=1,2,\cdots ,q$, $q<M$ as
\[\K_q = ^qY_f ^qY_p^\dagger\]
and use the corresponding $\phi_q$ and $z_q$ to compute the updated Koopman operators $\K_n$, $n>q$. However, one major issue of this approach is the invertibility of $\phi_q$, as for most practical purposes and applications, one would like $q$ to be small and this will imply that $\phi_q$ won't be of full rank, thus resulting in erroneous computation of the Koopman operator. To resolve this issue and to be more suitable for practical applications, we set 
\[\phi_0 = \delta I_K, \quad z_0 = 0_K,\]
where $\delta >0$, $I_K$ is the $K\times K$ identity matrix and $0_K$ is the $K\times K$ zero matrix.

\begin{remark}
Choosing the initialization parameter $\delta$ can be tricky and usually one should run the algorithm multiple times, with different $\delta$, on a given training data-set and choose the one which has lowest error on some validation data-set. 
\end{remark}

\begin{algorithm}[htp!]
\caption{Algorithm for online Koopman Operator computation using streaming data.}
\begin{enumerate}
\item{Fix the dictionary functions $\bf \Psi$.}
\item{Initialize $\phi_0 = \delta I_K$ and $ z_0 = 0_K$.}
\item{As a new data point $(x_{M},y_{M})$ streams in, lift the data point to $\mathbb{R}^K$ using the dictionary function $\bf \Psi$.}
\item{Update $z_{M}$ and $\phi_{M}^{-1}$ as
\begin{eqnarray*}
z_{M} &=& z_{M-1} + Y_f^{M}(Y_p^{M})^\top\\
\phi_{M}^{-1} &=& \phi_{M-1}^{-1} - \frac{\phi_{M-1}^{-1}Y_p^{M}(Y_p^{M})^\top\phi_{M-1}^{-1}}{1 + (Y_p^{M})^\top \phi_{M-1}^{-1} Y_p^{M}}.
\end{eqnarray*}
}
\item{Update the Koopman operator $\K_{M-1}$ to $\K_{M}$ as 
\begin{eqnarray*}
\K_{M} = z_{M}\phi_{M}^{-1}.
\end{eqnarray*}
}
\end{enumerate}
\label{algo}
\end{algorithm}

\section{Design of Koopman Predictor}\label{section_predictor}

 The main advantage of using Perron-Frobenius and Koopman operators for analysis of dynamical systems is the fact that these operators generate a linear system in a higher dimensional space, even if the underlying system is linear and the linearity of these operators can be used to design predictors for the underlying system. In this section, we briefly present the predictor design problem for the self-containment of the paper. For details we refer the readers to \cite{korda_mezic_predictor}. 

Let $\K_M$ be the Koopman operator computed from a streaming data-set $[x_1,\cdots , x_{M+1}]$, using algorithm \ref{algo} and let $\bar x_0$ be the initial condition from where the future trajectory needs to be predicted. Let  
 \[{\bf \Psi}(\bar x_0)=: {\bf z}_0\in \mathbb{R}^K\]
 be the data point in the lifted space. Then the initial condition is propagated using Koopman operator as \[{\bf z}_n={\bf K}_M^n{\bf z}_0.\] The predicted trajectory in the state space is then obtained as 
\[\bar x_n=C {\bf z}_n\]
where matrix $C$ is obtained as the solution of the following least squares problem
\begin{eqnarray}\label{C_pred}
\min_C\sum_{i = 1}^{M+1} \parallel x_i - C \boldsymbol \Psi (x_i)\parallel_2^2
\end{eqnarray}

\section{Simulation Results}\label{section_simulation}

In this section, we demonstrate the online computation of Koopman operator using streaming data for three different dynamical systems. Further, we compare the computation time of our proposed algorithm with the existing DMD and EDMD algorithms to illustrate the computational efficiency of the proposed recursive Koopman computation algorithm. All the simulations were performed in MATLAB\_R2018b on a Apple Macbook Pro with 2.3 GHz Intel Core i5 processor and 8 GB 2133 MHz LPDDR3 RAM. 

\subsection{Van der Pol Oscillator}
\begin{figure}[htp]
\centering
\includegraphics[scale=.3]{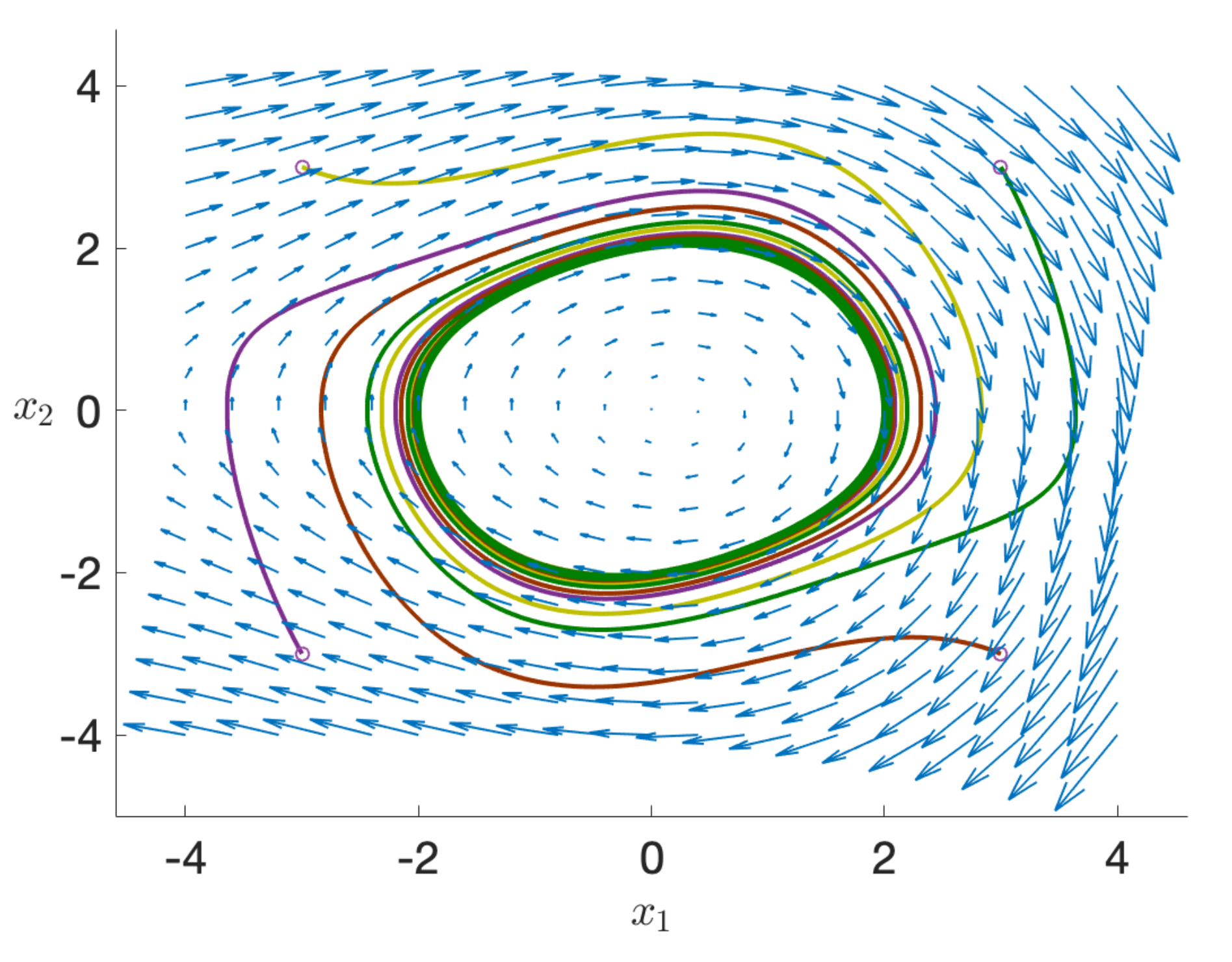}
\caption{Phase portrait of Van der Pol oscillator.}\label{vanderpol_phase}
\end{figure}

The Van der Pol system is a non-linear oscillator, with nonlinear damping and is given by 
\begin{eqnarray*}
\dot{x}_1 &=& x_2\\
\dot{x}_2 &=& \mu (1-x_1^2) - x_1
\end{eqnarray*}
where $\mu$ is the damping parameter. It can be seen that when $\mu=0$, the equations reduce to that of a harmonic oscillator, while for $\mu>0$, the system has a stable limit cycle. In this set of simulations, we choose $\mu=0.2$ and the corresponding phase portrait is shown in Fig. \ref{vanderpol_phase}. The stable limit cycle can be identified in Fig. \ref{vanderpol_phase}, as the trajectories converge there.

Since the dynamical system has one invariant set, the Koopman operator should have one eigenvalue equal to one and the corresponding eigenvector identifies the invariant set in the phase space\footnote{The system has an unstable equilibrium point at the origin and since it is of measure zero and only those initial points that start there remain there, the finite approximations of the transfer operators do not capture it.}. 

For simulation purposes, we chose one initial condition and propagated it in time and computed the Koopman operator iteratively using algorithm \ref{algo}, with the dictionary function set consisting of 60 Gaussian radial basis functions of the form $\psi(x) = \exp(-\parallel x \parallel^2/\sigma^2)$, where $\sigma=0.3$. In the simulations, we chose the initialization parameter $\delta=0.0001$. Intuitively, the initial Koopman operators, computed from a small data-set, should not be able to identify the invariant set. However, as the data set grows and the trajectory approaches the invariant set and is confined there, the Koopman operators should be able to identify the invariant set.

\begin{figure}[htp]
\centering
\subfigure[]{\includegraphics[scale=.214]{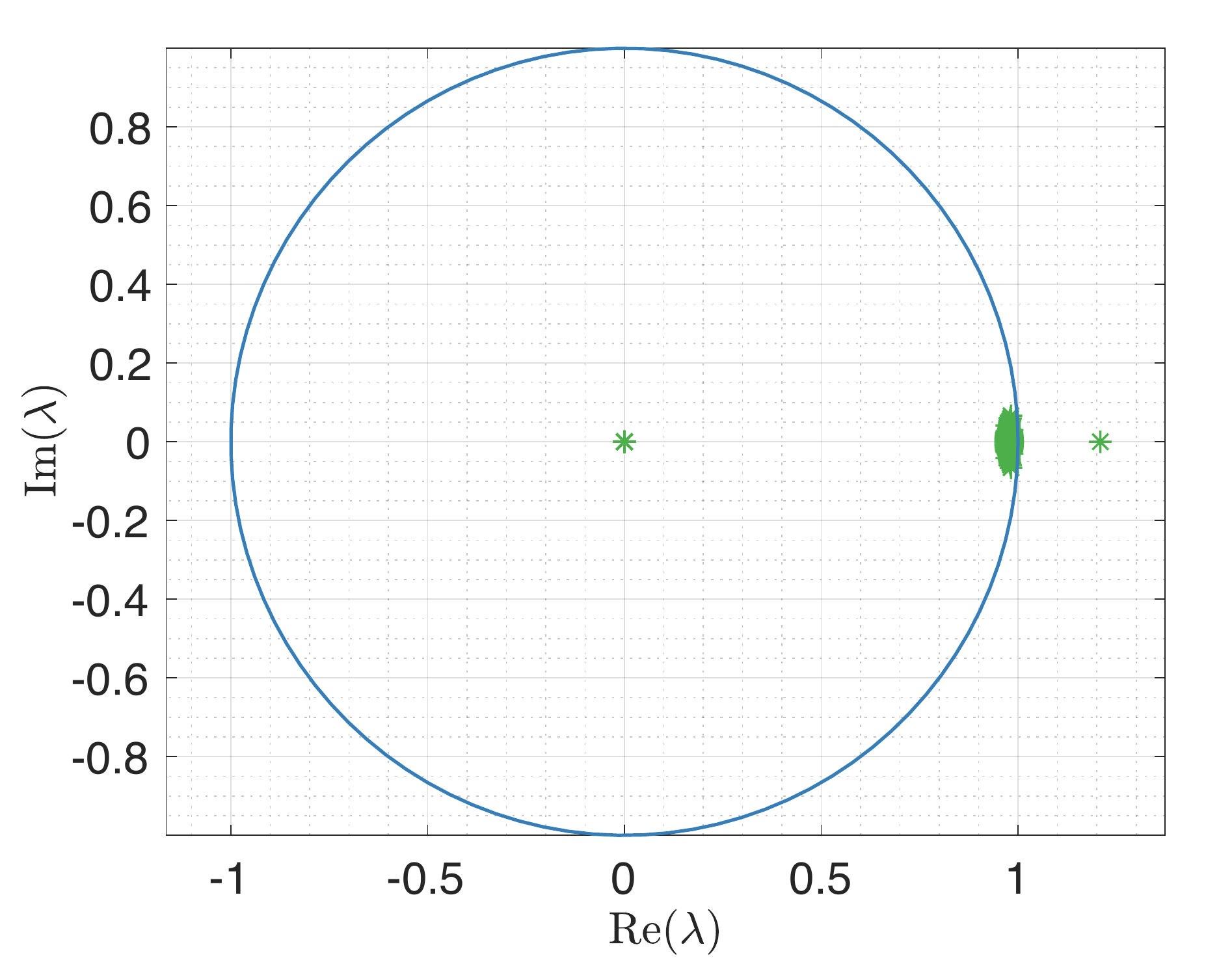}}
\subfigure[]{\includegraphics[scale=.21]{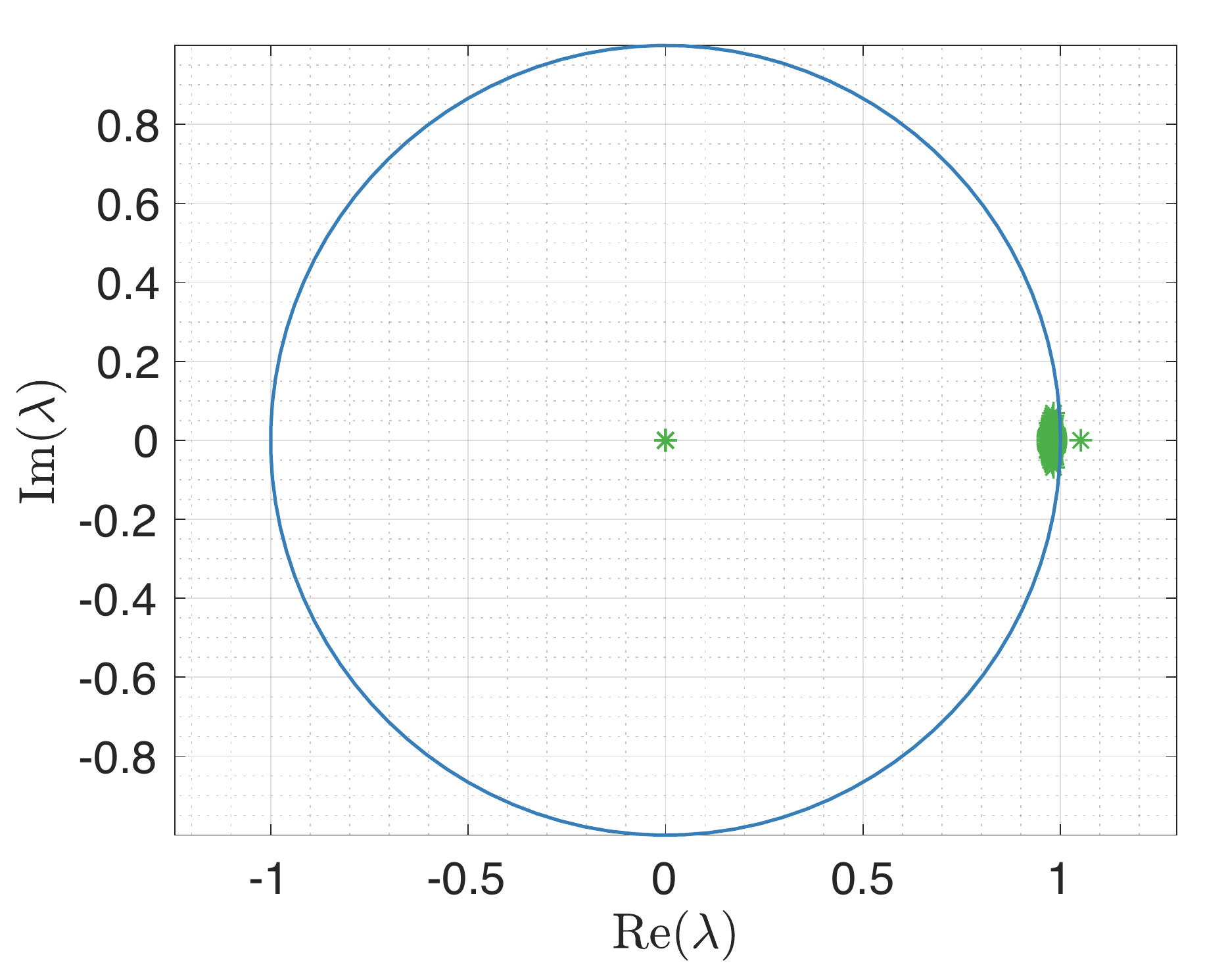}}
\subfigure[]{\includegraphics[scale=.21]{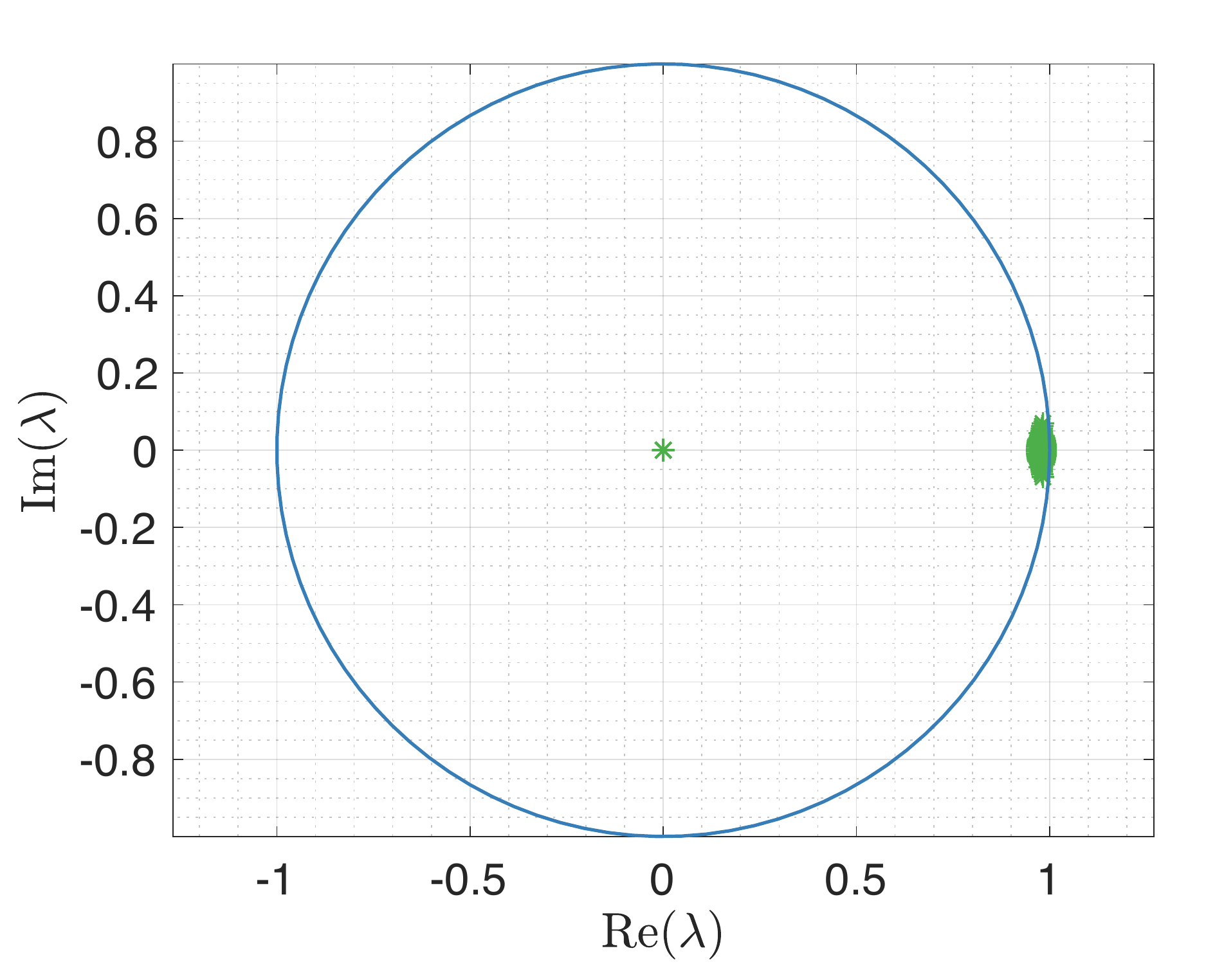}}
\caption{Eigenvalues of the recursive Koopman operator computed using (a) 1500 time steps, (b) 2000 time steps, (c) 2500 time steps data respectively}\label{vanderpol_eig}
\end{figure}
The Koopman operators were computed iteratively for 2500 time steps and the eigenvalues of the different Koopman operators are plotted in the complex plane in Fig. \ref{vanderpol_eig}. In \cite{sparse_Koopman_acc}, it was noted that if the data-set is not sufficiently large, the Koopman eigenvalues may be unstable. Figs. \ref{vanderpol_eig}(a) and (b) show similar results. In particular, the Koopman operators obtained using 1500 data points and 2000 data points are unstable, but the operator computed using 2500 data points is stable with one eigenvalue equal to one.

\begin{figure}[htp]
\centering
\subfigure[]{\includegraphics[scale=.21]{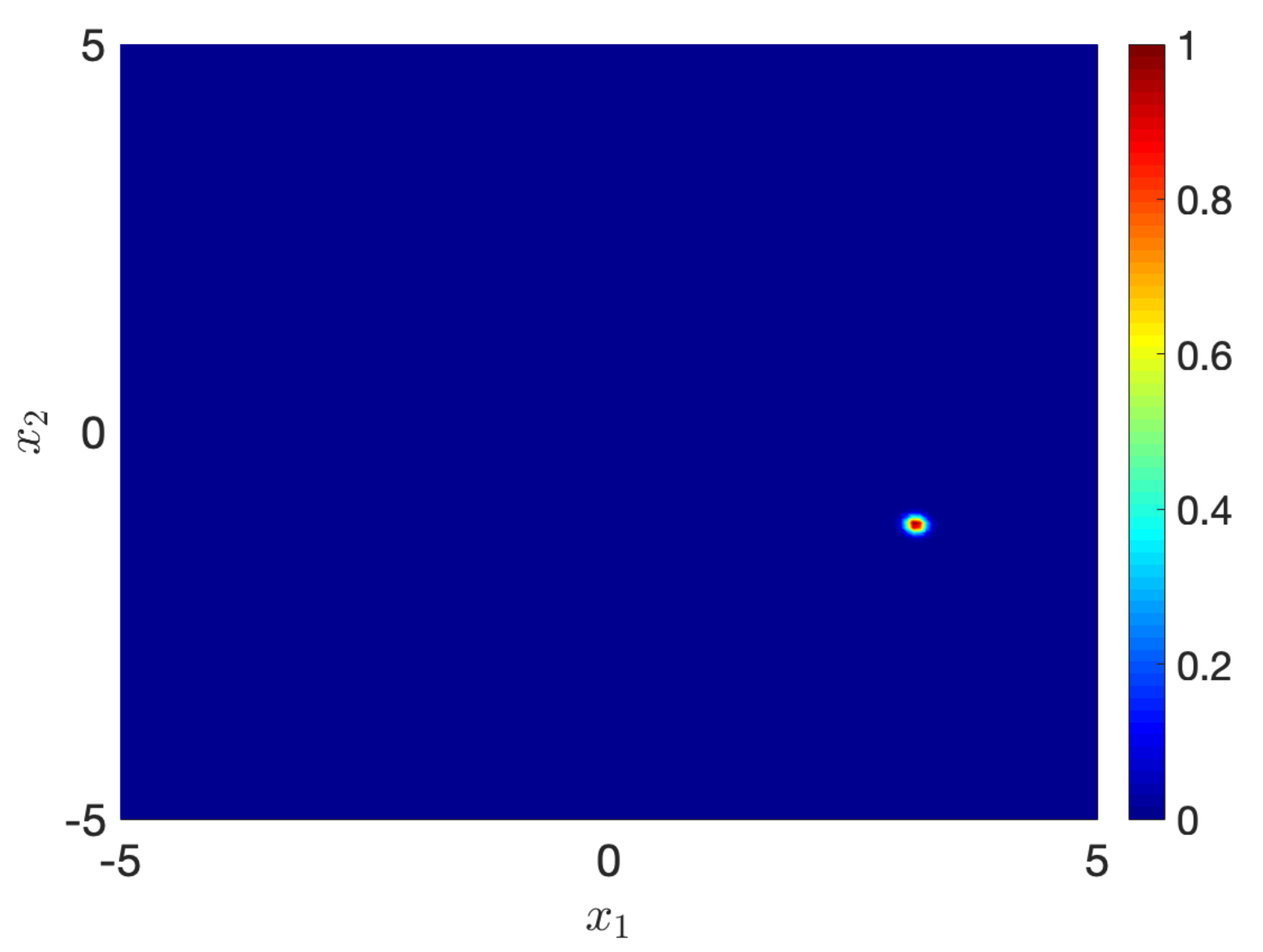}}
\subfigure[]{\includegraphics[scale=.21]{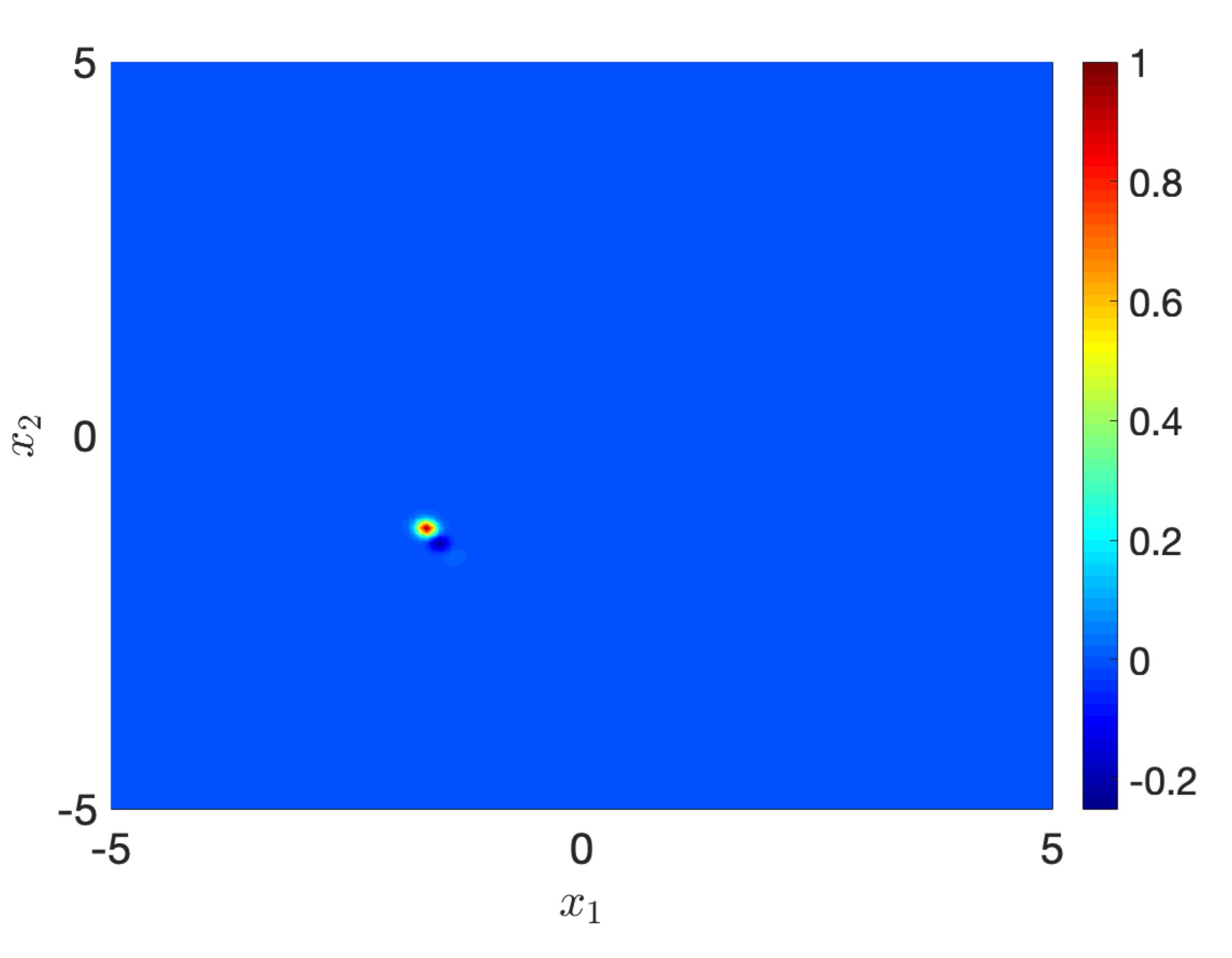}}
\subfigure[]{\includegraphics[scale=.21]{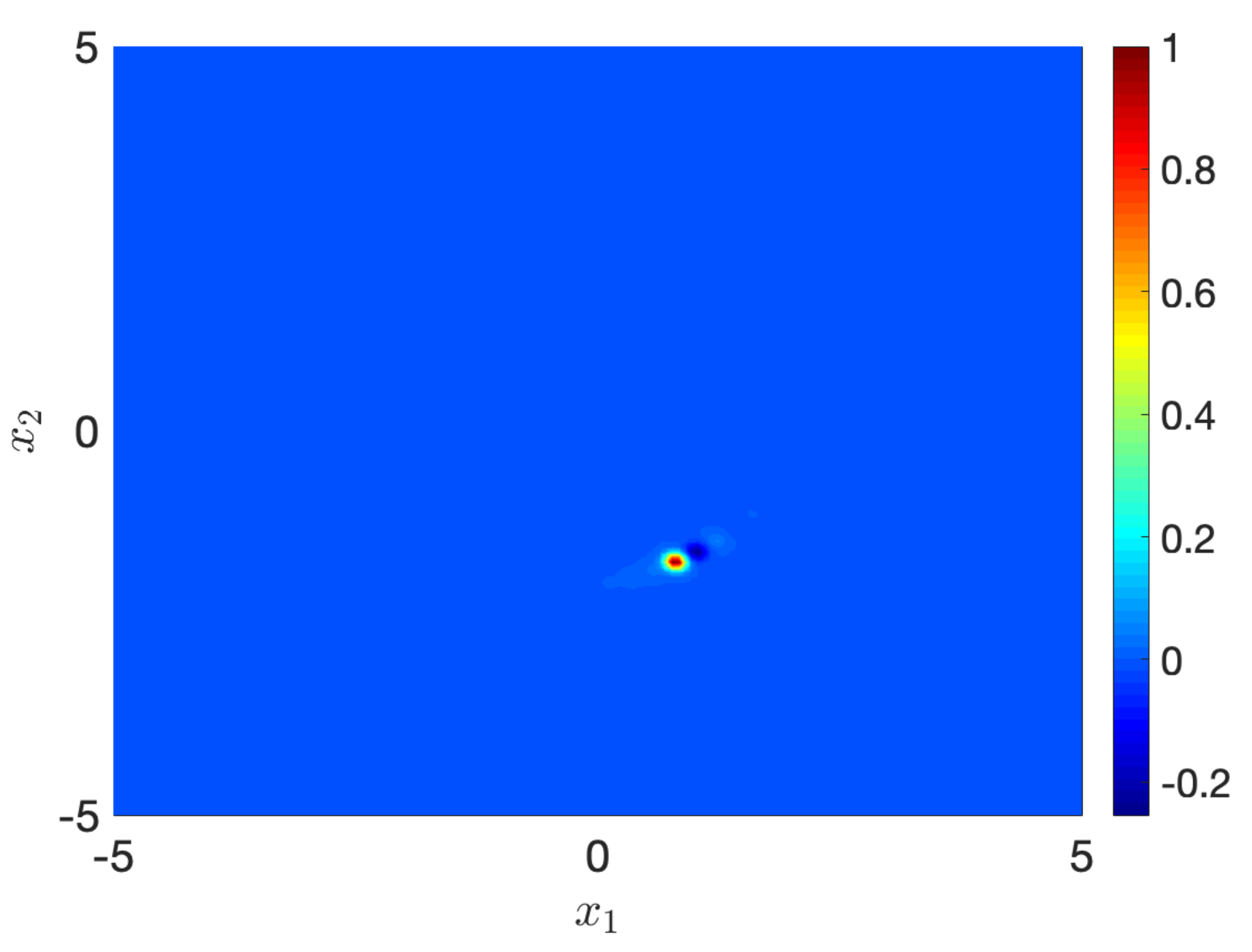}}
\subfigure[]{\includegraphics[scale=.21]{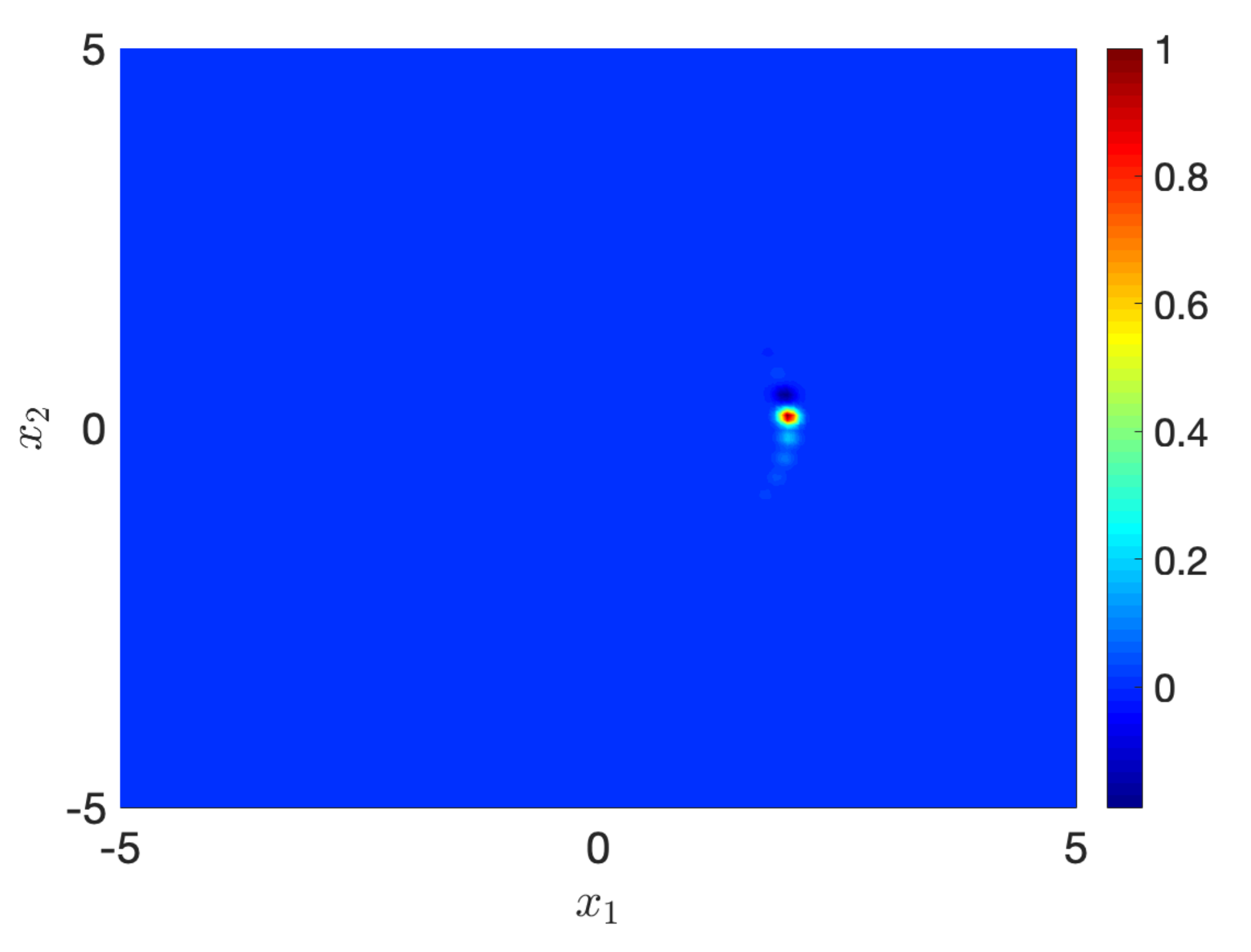}}
\subfigure[]{\includegraphics[scale=.21]{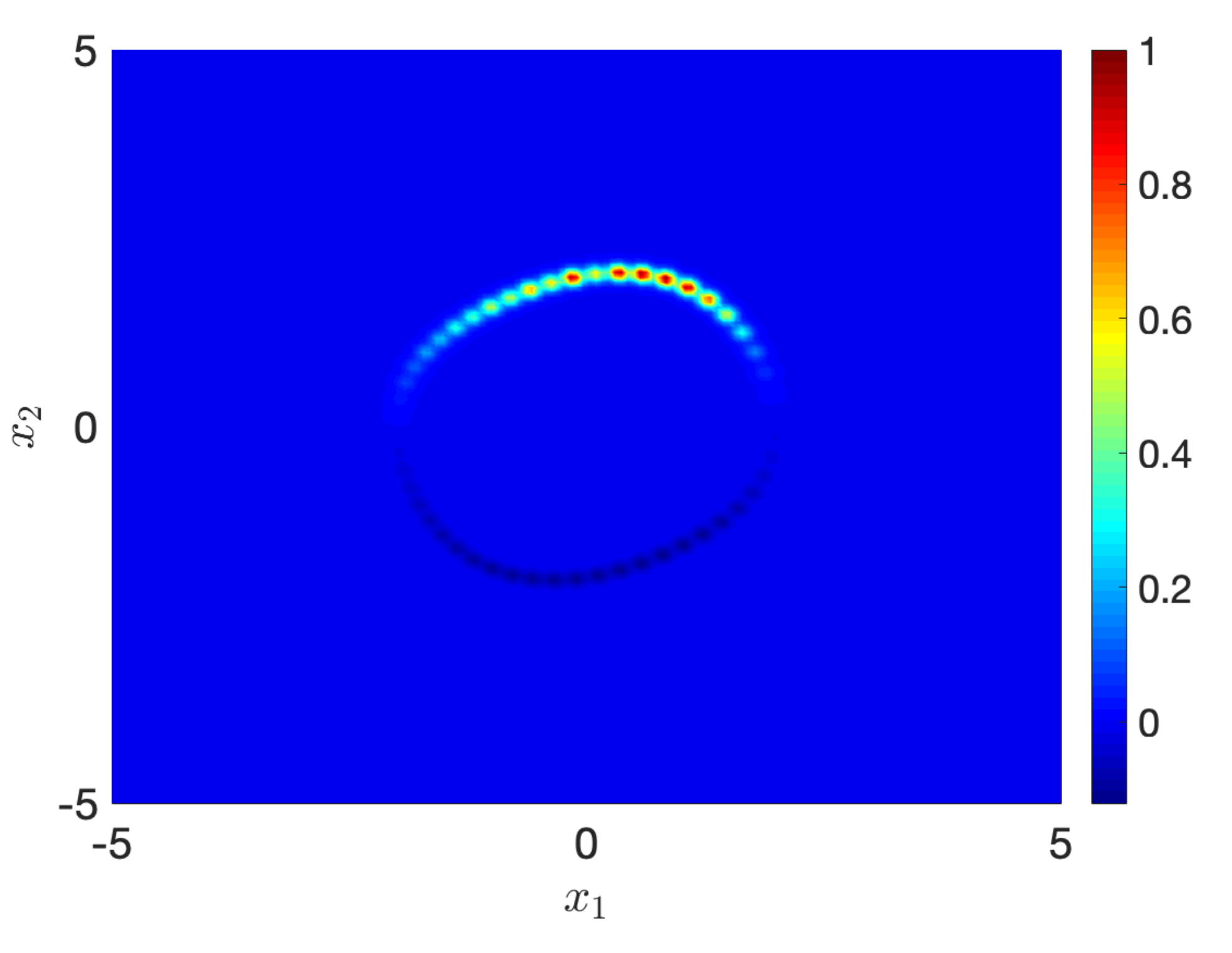}}
\caption{Koopman eigenfunction corresponding the largest eigenvalue of the Koopman operator. The figures correspond to the eigenfunctions computed iteratively using the first (a) 500 time steps, (b) 1000 time steps, (c) 1500 time steps, (d) 2000 time steps, and (e) 2500 time steps data respectively.}\label{inv_meas1}
\end{figure}

In theory, the eigenfunctions corresponding to unit eigenvalues corresponds to the invariant sets in the state space. However, in practice, since one works with finite-dimensional approximations, one usually looks at the eigenfunctions which correspond to eigenvalues with largest real parts. In Fig. \ref{inv_meas1} the eigenfuctions corresponding to the largest eigenvalue of the computed Koopman operators are plotted. It can be observed that the recursive algorithm gradually recovers the invariant set, as the size of the data set is increased. 

\begin{figure}[htp]
\centering
\subfigure[]{\includegraphics[scale=.21]{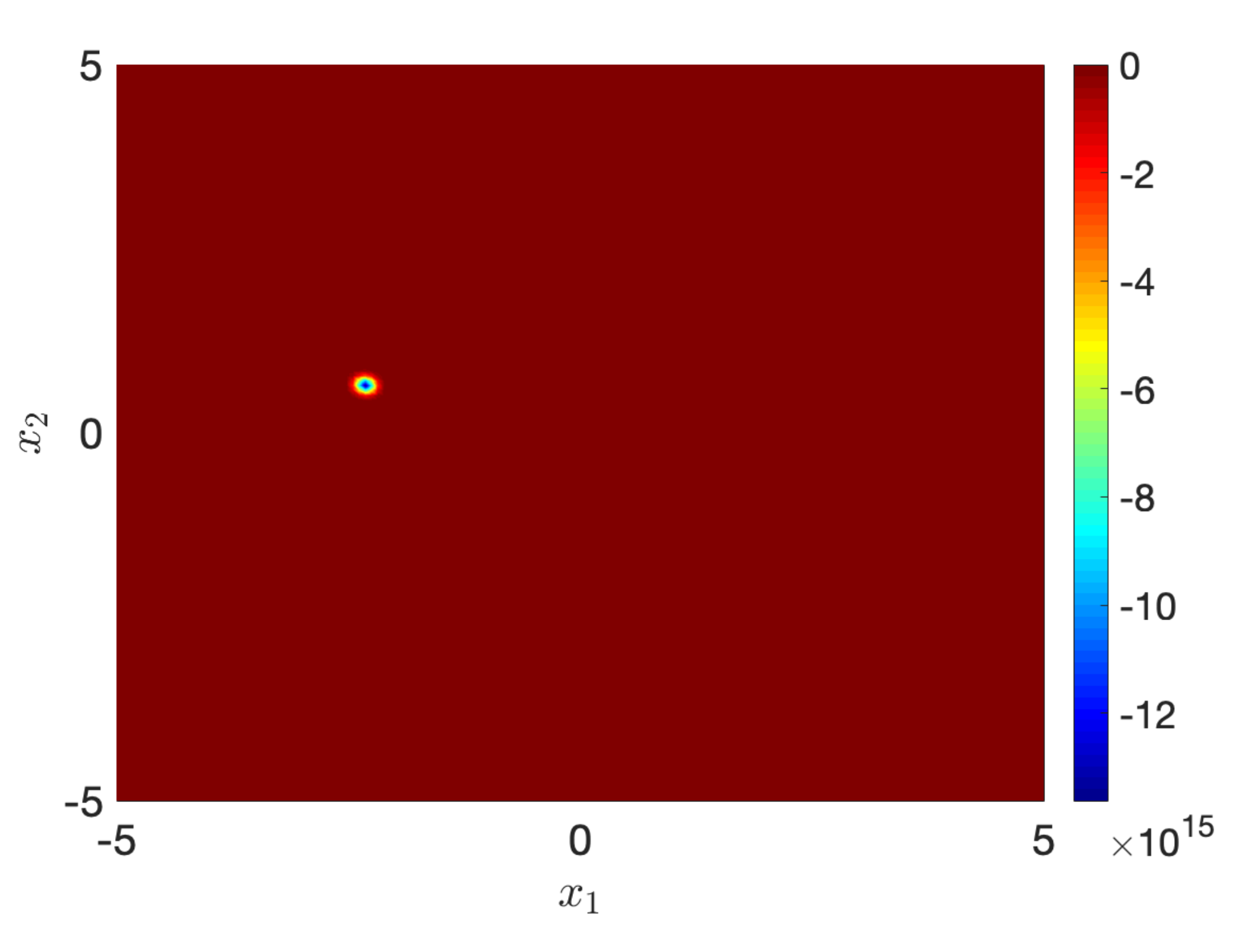}}
\subfigure[]{\includegraphics[scale=.21]{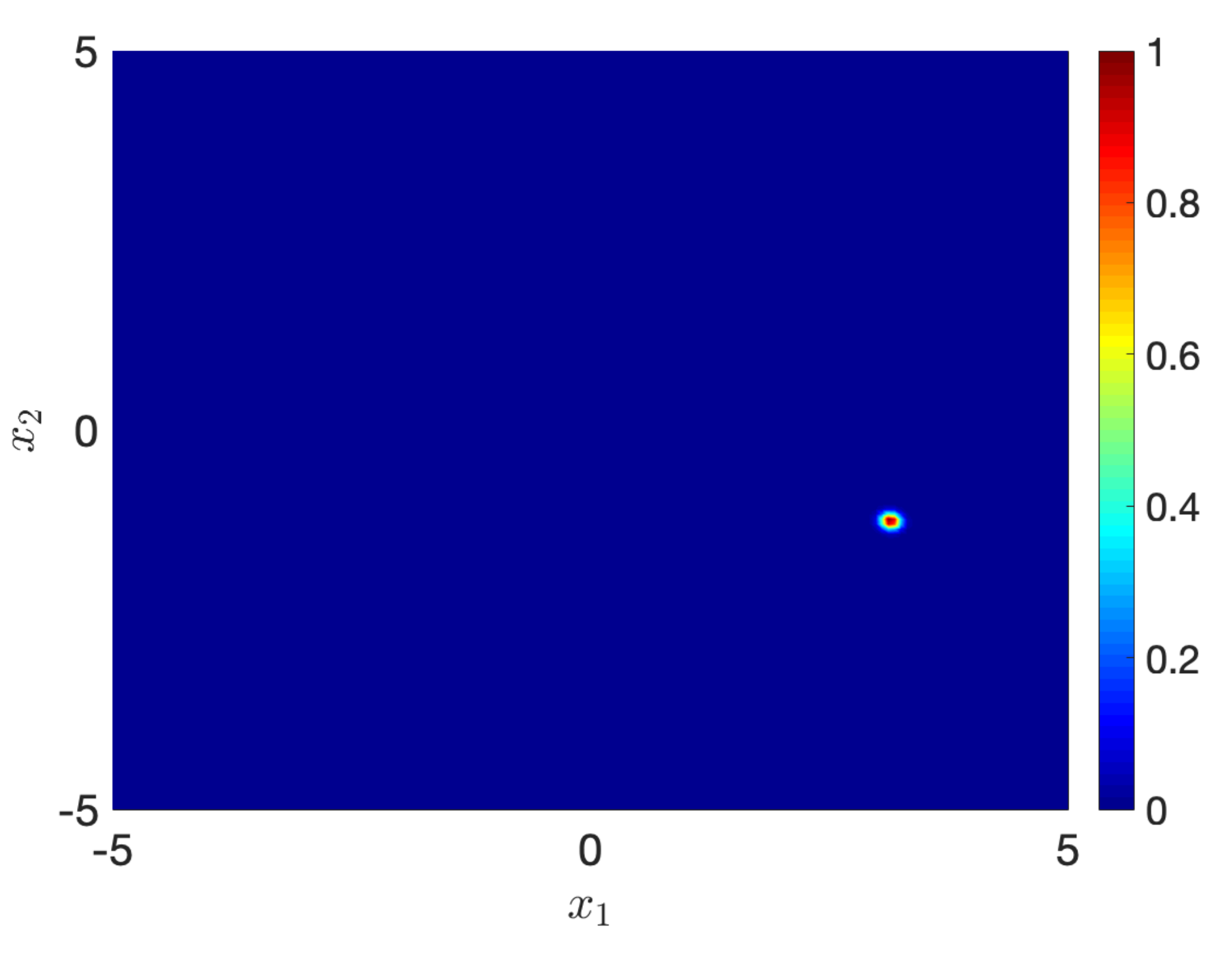}}
\subfigure[]{\includegraphics[scale=.21]{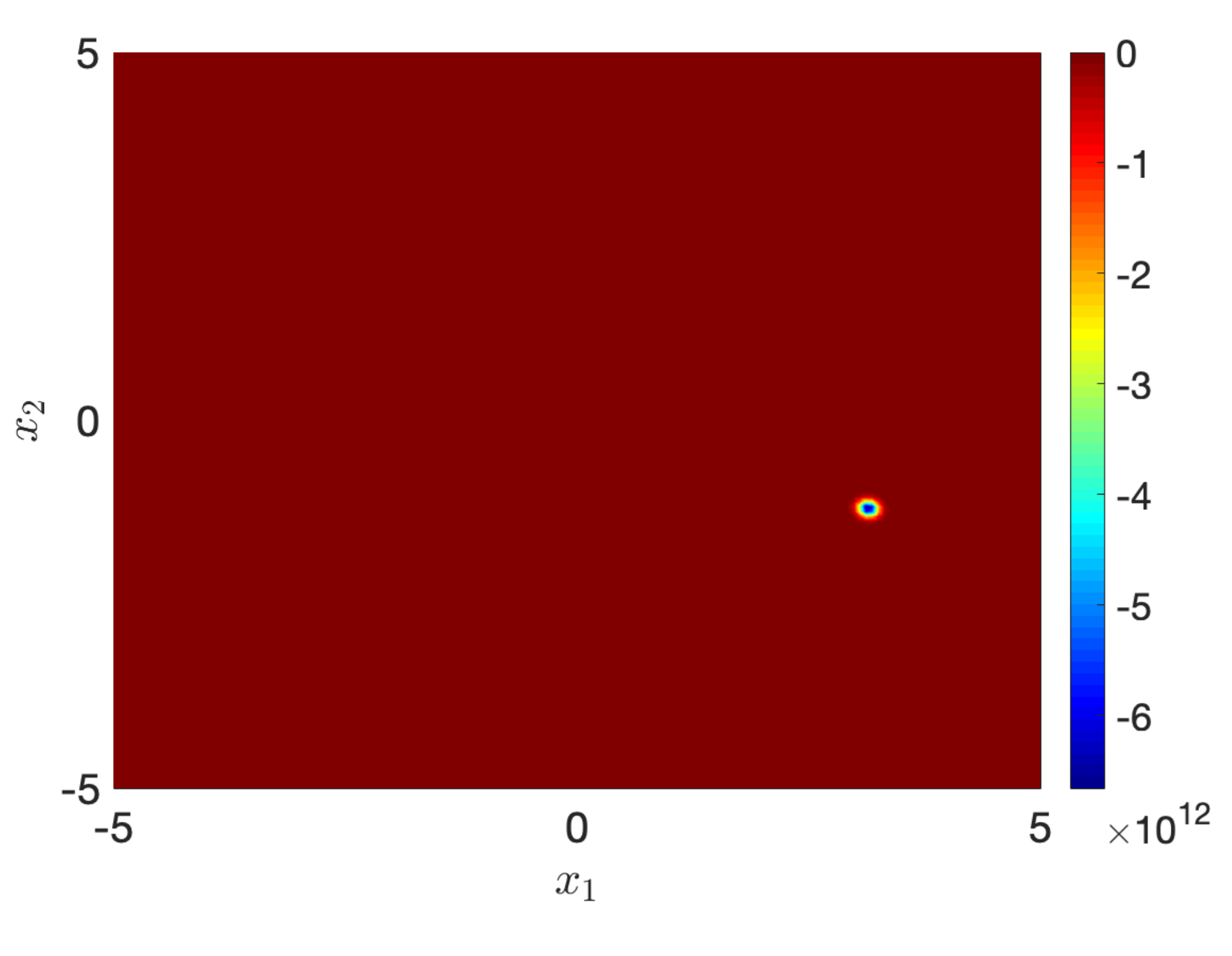}}
\subfigure[]{\includegraphics[scale=.21]{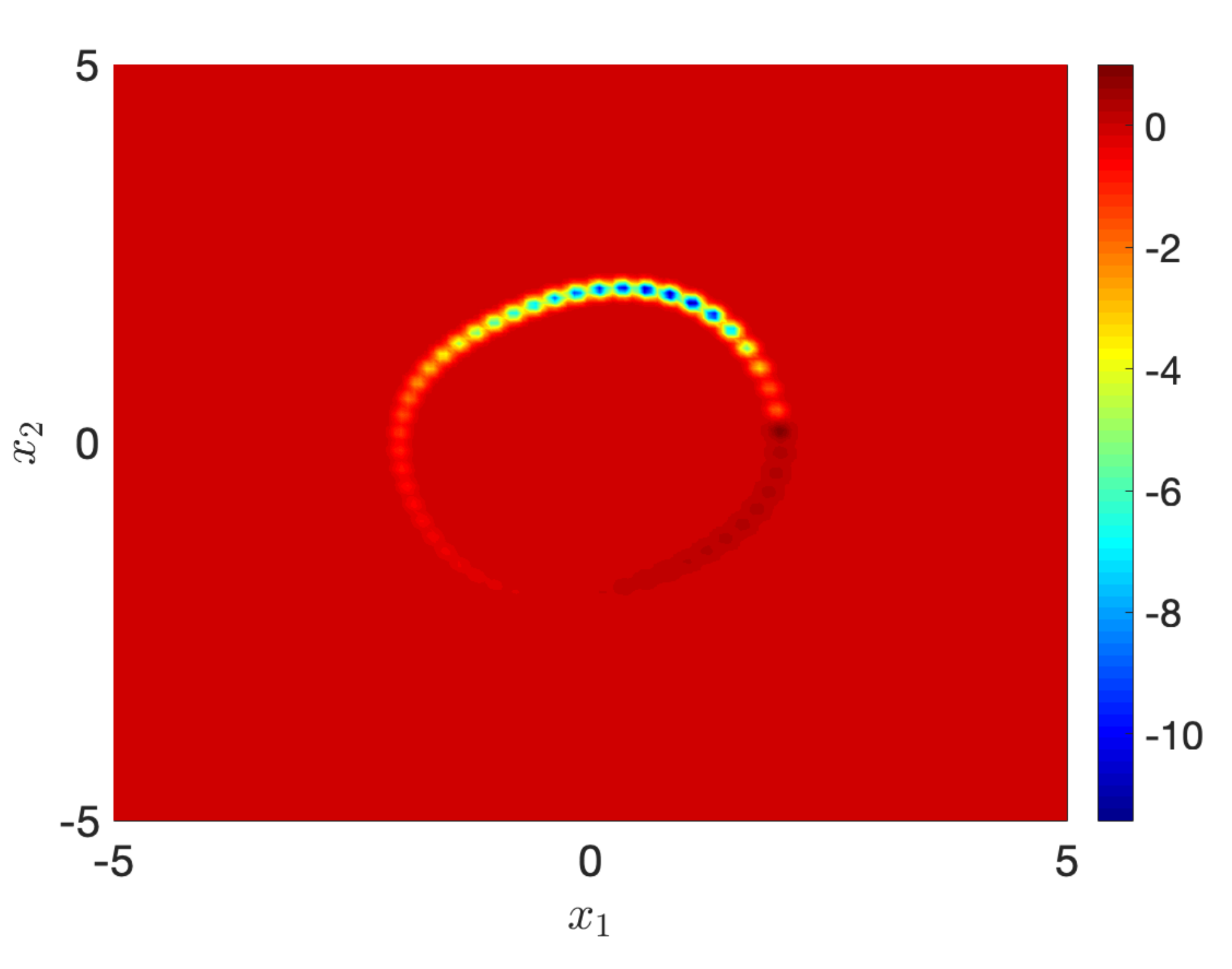}}
\subfigure[]{\includegraphics[scale=.21]{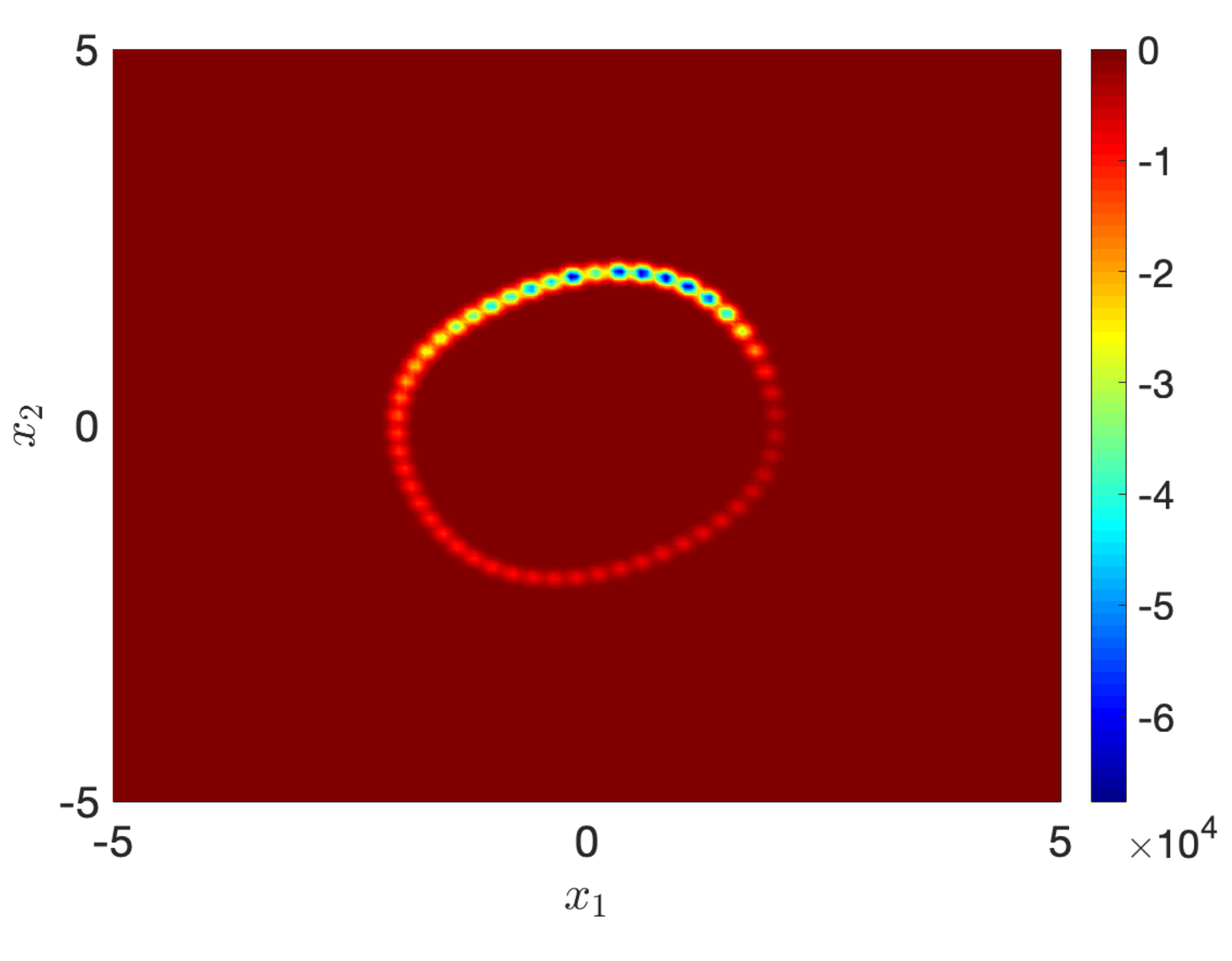}}
\caption{Koopman eigenfunction corresponding the second largest eigenvalue of the Koopman operator. The figures correspond to the eigenfunctions computed iteratively using the first (a) 500 time steps, (b) 1000 time steps, (c) 1500 time steps, (d) 2000 time steps, and (e) 2500 time steps data respectively.}\label{inv_meas2}
\end{figure}

Fig. \ref{inv_meas2} shows the eigenfunctions corresponding to the second largest eigenvalue. Since the second largest eigenvalue is also close to one, these eigenfunctions also identify the invariant set. These plots shows the efficiency of our iterative algorithm. However, the main advantage of the iterative Koopman computation algorithm lies in the computation time.
{\small
\begin{table}[htp!]
\centering
\caption{Comparison of computation time of Recursive Koopman Operator and EDMD}\label{vanderpol_comp_time}
\begin{tabular}{|c|c|c|c|}
\hline
\# of Data points/ & \multicolumn{1}{c|}{\multirow{2}{*}{Iterative Koopman Operator}} & \multirow{2}{*}{DMD} \\
Computation time by  & \multicolumn{1}{c|}{} &  \\ 
\hline
    $1500$ & $0.921$s & $4.411$s \\
     $2000$ & $1.292$s & $7.703$s\\
     $2500$ & $1.728$s  & $11.199$s\\
\hline
\end{tabular}
\end{table}
}

In Table \ref{vanderpol_comp_time} we compare the computation time of the Koopman operator and the Koopman eigenvalues by the recursive algorithm \ref{algo} and by EDMD, using streaming data. In particular, we use our proposed algorithm to compute the Koopman operator and its eigenvalues at each time step till 1500, 2000 and 2500 time steps and compare the time required to compute the same using EDMD algorithm. In other words, in our proposed method, we update the Koopman operator iteratively as a new data point streams in and in the usual EDMD approach, every time a new data point streams in, the Koopman operator is computed from scratch, without using any information from the Koopman operator computed at the previous time step. From Table \ref{vanderpol_comp_time} it can be clearly seen that the proposed algorithm computes the Koopman operator and its eigenvalues much more efficiently as compared to the existing EDMD algorithm.

\subsection{Network of Coupled Oscillators}

\begin{figure}[htp]
\centering
\subfigure[]{\includegraphics[scale=.3]{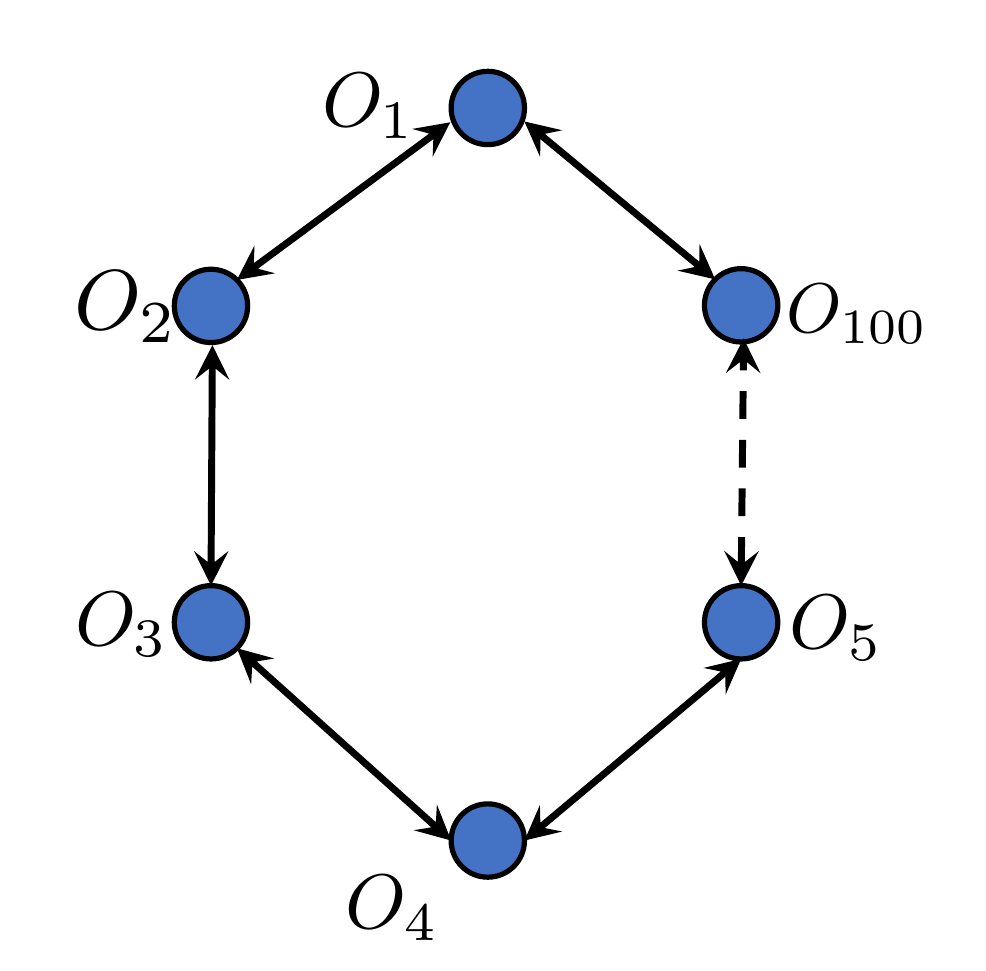}}
\subfigure[]{\includegraphics[scale=.19]{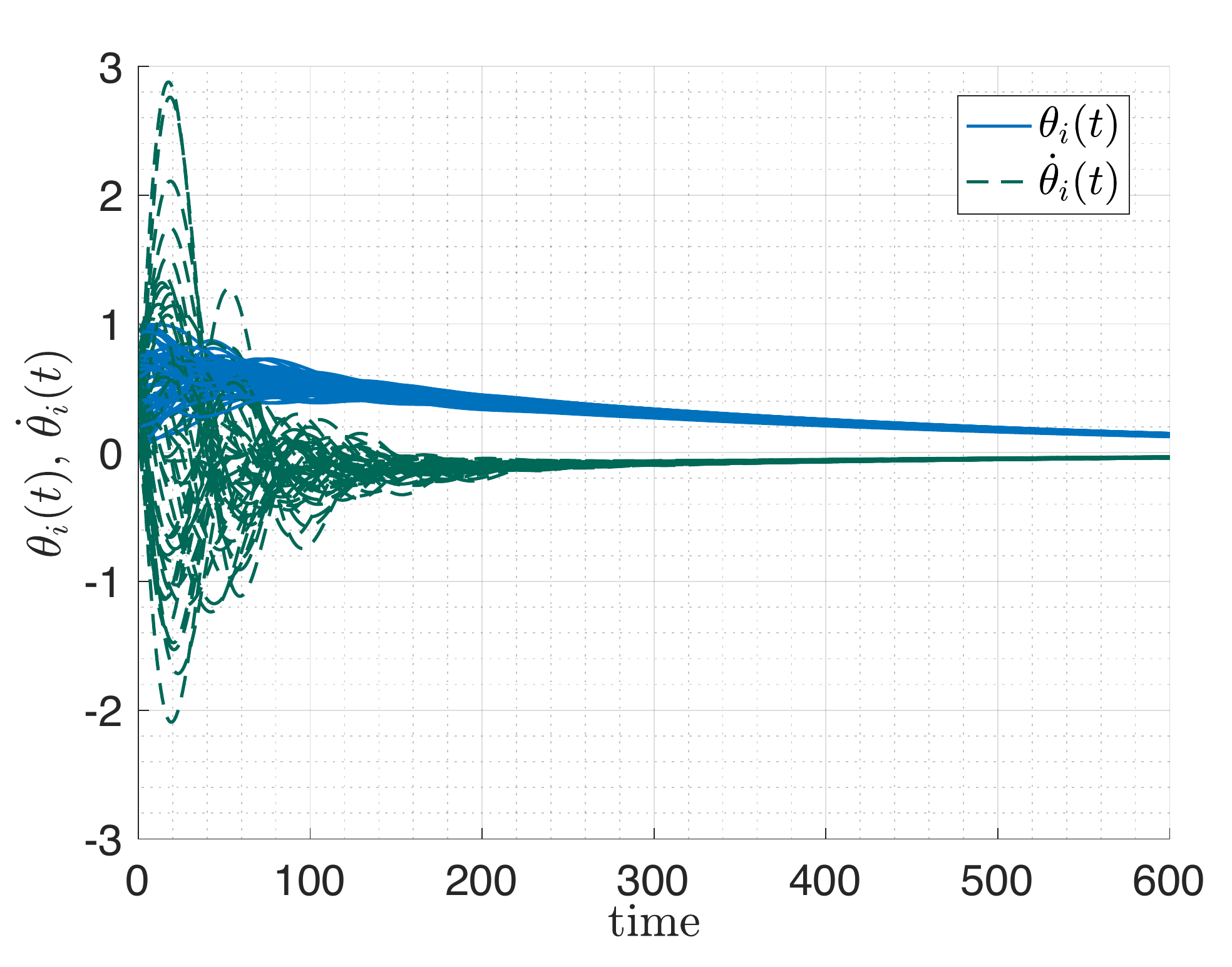}}
\caption{(a) Ring network of oscillators. (b) Time domain trajectories of the oscillators.}\label{oscillators}
\end{figure}

Consider a network of coupled linear oscillators given by
\begin{eqnarray}\label{coup_osc}
\ddot{\theta}_k &=& -\mathcal{L}_k\theta - d\dot{\theta}_k, \quad k = 1,\cdots , N
\end{eqnarray}
where $\theta_k$ is the angular position of the $k^{th}$ oscillator, $N$ is the number of oscillators, $\mathcal{L}_k$ is the $k^{th}$ row of the Laplacian $\mathcal{L}$ and $d$ is the damping coefficient. The Laplacian $\cal L$ is chosen such that the network is a ring network with 100 oscillators, as shown in Fig. \ref{oscillators}(a). 

For simulation purposes, we chose one initial condition and set the damping coefficient $d=2$ and the corresponding time domain trajectories are shown in Fig. \ref{oscillators}(b). 

\begin{figure}[htp]
\centering
\subfigure[]{\includegraphics[scale=.2]{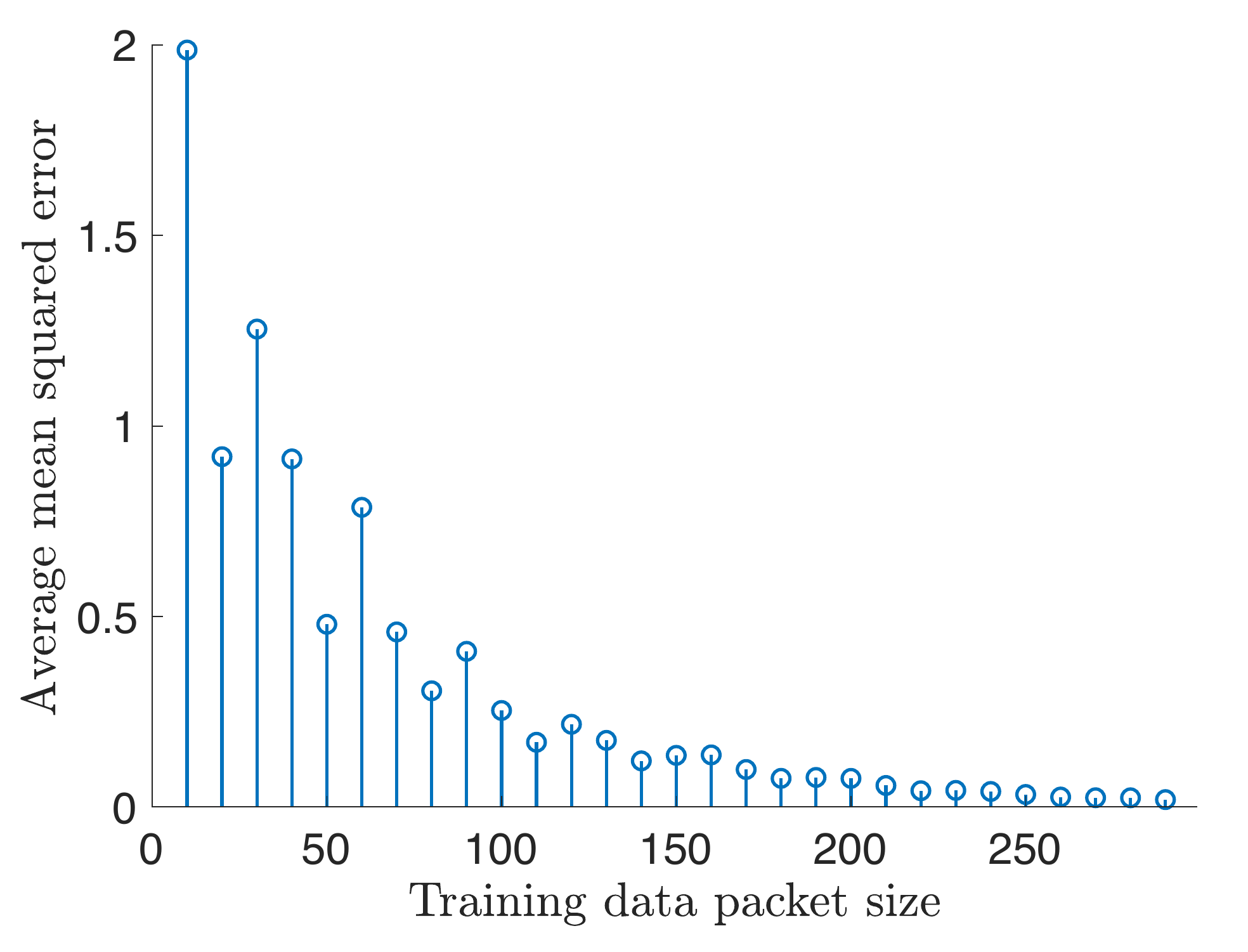}}
\subfigure[]{\includegraphics[scale=.2]{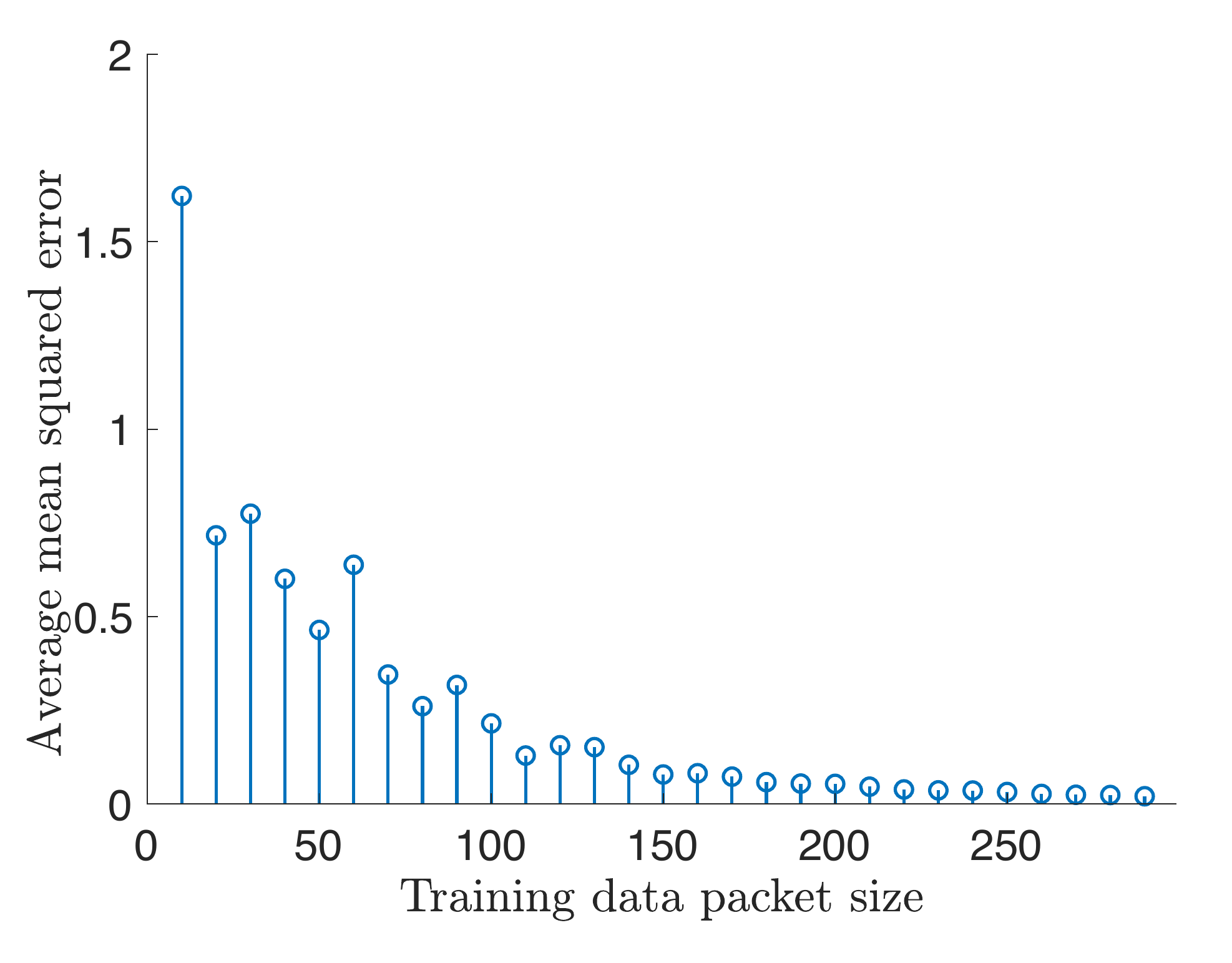}}
\caption{(a) Average mean squared error in prediction of the states of immediate 50 time steps using the incrementally computed Koopman operator till $t=300$. (b) Average mean squared error in prediction of the states for 50 time steps from time step $t=401$ to $t=450$ using incremental training data size till $t=300$.}\label{osc_pred}
\end{figure}

In this example, we used linear dictionary functions and computed the Koopman operator incrementally till $300$ time steps, with the initialization parameter $\delta=0.0001$. Further, using the predictor formulation of \cite{korda_mezic_predictor}, we predict the future of evolution of the states. The average mean squared errors in prediction of all the states, as the Koopman operator is updated, is shown in Fig. \ref{osc_pred}. In Fig. \ref{osc_pred}(a) we predict the future evolution of the states from time step $t_0+1$ to $t_0+50$ using the Koopman operator computed using the data upto time step $t=t_0$. As expected, the average error in prediction goes down as the training size data increases.

Note that, in this case, the predictor starts predicting the future with the data at time $t_0$ as the initial condition using the Koopman operator computed with data till time $t=t_0$. Hence, the training data and the test data has one time point data common. To test how our algorithm works on a test data-set it has not seen, we tried to predict the evolution of the states from time $t=401$ to time $t=450$. The corresponding average mean squared errors are shown in Fig. \ref{osc_pred}(b). It can be observed that the Koopman operator computed by our proposed iterative algorithm, coupled with the predictor, performs really well on the unseen data-set.

{\small
\begin{table}[htp!]
\centering
\caption{Comparison of computation time of Recursive Koopman Operator and DMD}\label{osc_comp_time}
\begin{tabular}{|c|c|c|c|c|}
\hline
\# of Data points/ & \multicolumn{1}{c|}{\multirow{2}{*}{Iterative Koopman Operator}} & \multirow{2}{*}{DMD} \\
Computation time by  & \multicolumn{1}{c|}{} &  \\ 
\hline
    $200$ & $0.1644$s & $0.2847$s \\
     $250$ & $0.2075$s & $0.4530$s\\
     $300$ & $0.2428$s  & $0.7541$s\\
\hline
\end{tabular}
\end{table}
}

Further, in Table \ref{osc_comp_time}, we compare the online computation time of the Koopman operator using our proposed iterative approach and the normal DMD algorithm, using streaming data. As in the Van der Pol oscillator example, we see that our proposed algorithm is more efficient than the existing DMD algorithm.

\subsection{Burger Equation}

The third example considered in this paper is the Burger equation. Burger equation is a successful but simplified partial differential equation which describes the motion of viscous compressible fluids. The equation is of the form 
\[\partial_t u(x,t)+u\partial_x u=k\partial_x^2u\]
where $u$ is the speed of the gas, $k$ is the kinematic viscosity, $x$ is the spatial coordinate and $t$ is time. 
\begin{figure}[htp]
\centering
\includegraphics[scale=.35]{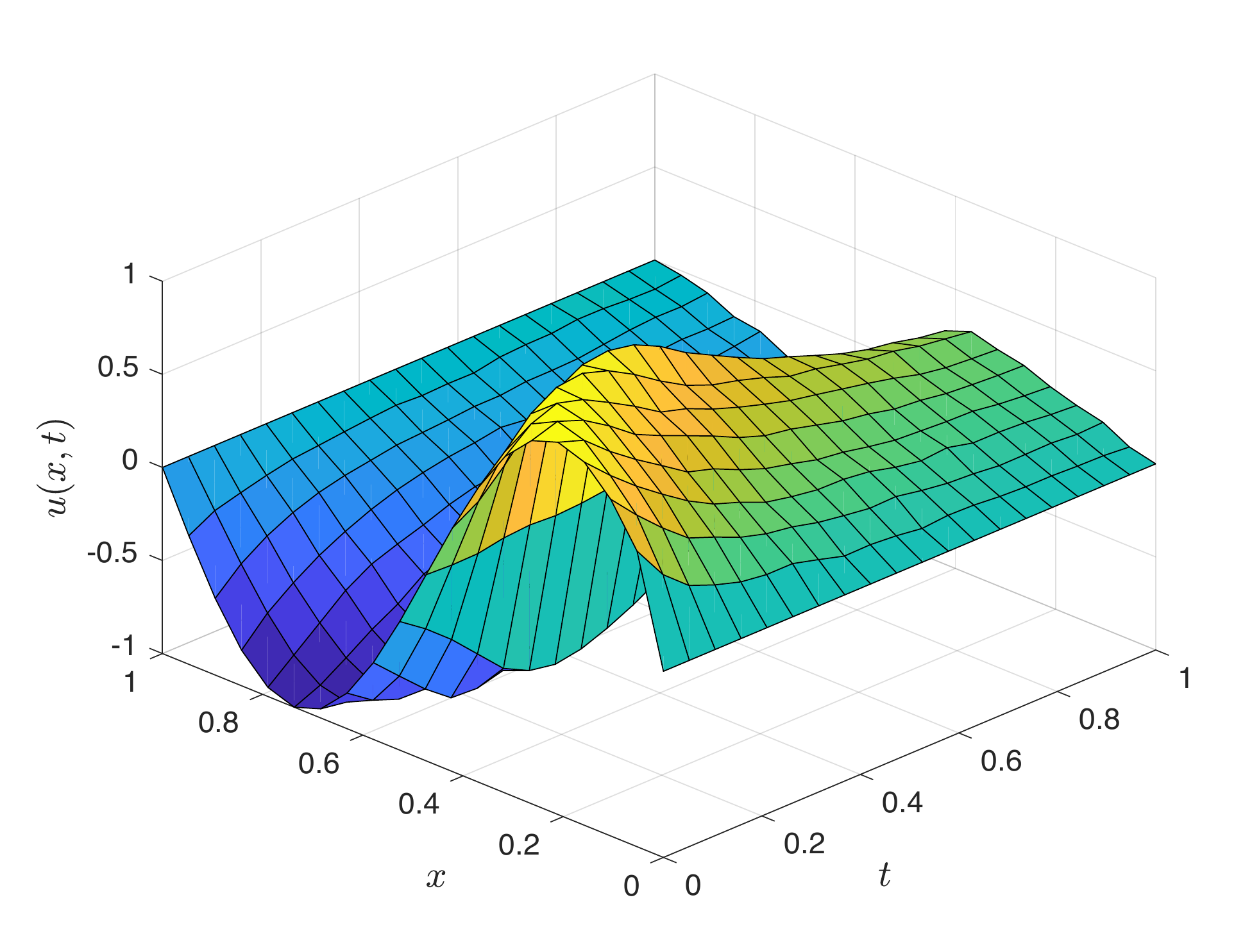}
\caption{Flow field of Burger equation.}\label{burger_flow}
\end{figure}

In the simulation, choosing $k=0.01$, we approximated the PDE solution using the Finite Difference method \cite{KUTLUAY1999251} with the initial condition $u(x,0)=sin(2\pi x)$ and Dirichet boundary condition $u(0,t)=u(1,t)=0$. Given the spatial and temporal ranges, $x\in[0,1],\;t\in[0,1]$, the discretizaion steps are chosen as $\Delta t=0.02$ and $\Delta x=1\times10^{-2}$. With the above set of conditions, the flow $u$ is shown in Fig. \ref{burger_flow}.

\begin{figure}[htp]
\centering
\subfigure[]{\includegraphics[scale=.2]{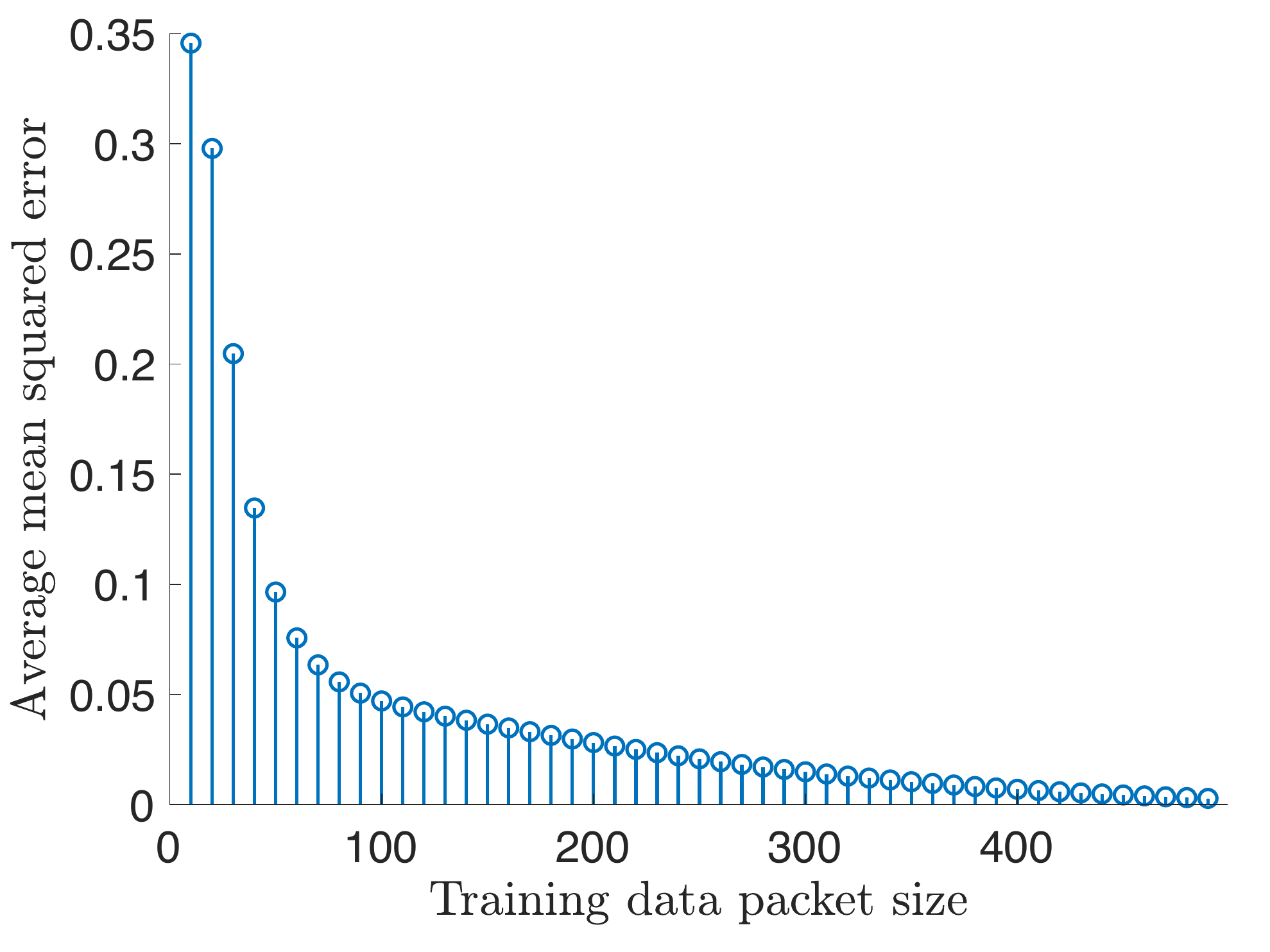}}
\subfigure[]{\includegraphics[scale=.2]{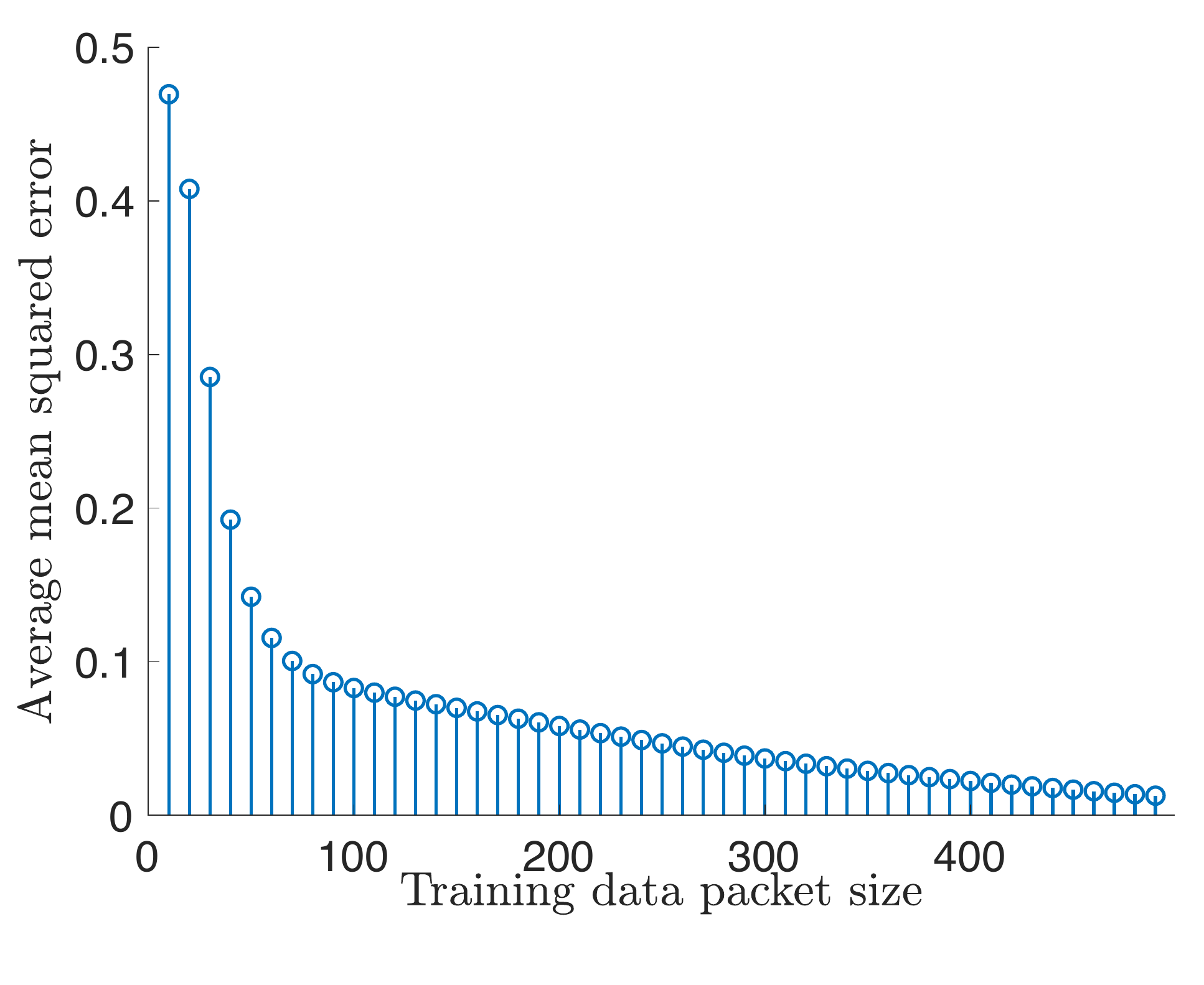}}
\caption{(a) Average mean squared error in prediction of the states of immediate 50 time steps using the incrementally computed Koopman operator till $t=500$. (b) Average mean squared error in prediction of the states for 50 time steps from time step $t=701$ to $t=750$ using incremental training data size till $t=500$.}\label{burger_pred}
\end{figure}

Since the space discretization was chosen as $\Delta x = 1\times 10^{-2}$, there are 100 state variables. In this example we used linear dictionary functions for the computation of the Koopman operator. In particular, we used data points till $t=500$ time steps to compute the recursive Koopman operator at each time step, using streaming data. The initialization parameter was chosen to be $\delta=0.0001$. The recursive Koopman operator computed was further used to predict the future evolution of the states. As in oscillator example, we predict the future in two different cases. In one instance, as the Koopman operated is computed at each time step, it was used the predict the immediate future 50 time step trajectories. The average mean squared error in prediction of the states is shown in Fig. \ref{burger_pred}(a). In the second case, the Koopman operators computed at each time step was used to predict the evolution of trajectories from time step $t=701$ to time step $t=750$. This was done to analyze how the Koopman operators perform on an unseen data set. The average mean squared errors are shown in Fig. \ref{burger_pred}(b).

Further, we compare the computation time of the Koopman operator computation using the proposed algorithm and existing DMD algorithm and the results are shown in Table \ref{burger_comp_time}. As in the previous cases, we find that the proposed algorithm is computationally much more efficient than the DMD algorithm for computing the Koopman operator with streaming data. 


\begin{table}[]
\caption{Comparison of computation time of Recursive Koopman Operator and DMD}\label{burger_comp_time}
\begin{center}
\begin{tabular}{|c|c|c|}
\hline
\# of Data points/ & \multicolumn{1}{c|}{\multirow{2}{*}{Iterative Koopman Operator}} & \multirow{2}{*}{DMD} \\
Computation time by  & \multicolumn{1}{c|}{} &  \\ 
\hline
    $350$ & $0.0464$s & $0.4404$s \\
    $400$ & $0.0506$s & $0.5167$s \\
     $450$ & $0.0605$s & $0.6375$s\\
     $500$ & $0.0662$s  & $0.7762$s\\
\hline
\end{tabular}
\end{center}
\end{table}

\section{Conclusions}\label{section_conclusion}

Koopman operator theoretic analysis of dynamical systems has gained immense attention in recent years as it facilitates data-driven analysis of dynamical systems. Again, in many applications like power networks, biological systems, building systems, it is often necessary to analyze the system in real time. To this end, it is required to update the Koopman operator as new data points stream in. Existing algorithms for computing the Koopman operator uses batch data and the algorithms are not iterative in the sense that if a new data point streams in, the Koopman operator needs to computed from scratch using the entire data set. In this paper, we proposed an iterative algorithm for computation of the Koopman operator, such that the Koopman operator obtained at some particular time step is updated incrementally as the next snapshot of data arrives, thus doing away with the Koopman computation from scratch. We demonstrated the efficiency of our algorithm by constructing the Koopman operator for three different systems and also illustrate the computational superiority (with respect to time) of the proposed algorithm, compared to the existing DMD and EDMD algorithms.

\section{Acknowledgement}
The authors would also like to thank Umesh Vaidya, Nathan Kutz, Nibodh Boddupalli, and Igor Mezic for stimulating conversations and feedback. Any opinions, findings and conclusions or recommendations expressed in this material are those of the author(s) and do not necessarily reflect the views of the Defense Advanced Research Projects Agency (DARPA), the Department of Defense, or the United States Government. This work was supported partially by a Defense Advanced Research Projects Agency (DARPA) Grant No. DEAC0576RL01830 and an Institute of Collaborative Biotechnologies Grant.

\bibliographystyle{IEEEtran}
\bibliography{subhrajit_robust_DMD}

\begin{thebibliography}{10}
\providecommand{\url}[1]{#1}
\csname url@samestyle\endcsname
\providecommand{\newblock}{\relax}
\providecommand{\bibinfo}[2]{#2}
\providecommand{\BIBentrySTDinterwordspacing}{\spaceskip=0pt\relax}
\providecommand{\BIBentryALTinterwordstretchfactor}{4}
\providecommand{\BIBentryALTinterwordspacing}{\spaceskip=\fontdimen2\font plus
\BIBentryALTinterwordstretchfactor\fontdimen3\font minus
  \fontdimen4\font\relax}
\providecommand{\BIBforeignlanguage}[2]{{%
\expandafter\ifx\csname l@#1\endcsname\relax
\typeout{** WARNING: IEEEtran.bst: No hyphenation pattern has been}%
\typeout{** loaded for the language `#1'. Using the pattern for}%
\typeout{** the default language instead.}%
\else
\language=\csname l@#1\endcsname
\fi
#2}}
\providecommand{\BIBdecl}{\relax}
\BIBdecl

\bibitem{Dellnitz_Junge}
M.~Dellnitz and O.~Junge, ``On the approximation of complicated dynamical
  behavior,'' \emph{SIAM Journal on Numerical Analysis}, vol.~36, pp. 491--515,
  1999.

\bibitem{Mezic2000}
I.~Mezic and A.~Banaszuk, ``Comparison of systems with complex behavior:
  spectral methods,'' in \emph{Proceedings of the 39th IEEE Conference on
  Decision and Control (Cat. No.00CH37187)}, vol.~2, 2000, pp. 1224--1231
  vol.2.

\bibitem{froyland_extracting}
G.~Froyland, ``Extracting dynamical behaviour via {Markov} models,'' in
  \emph{Nonlinear Dynamics and Statistics: Proceedings, Newton Institute,
  Cambridge, 1998}, A.~Mees, Ed.\hskip 1em plus 0.5em minus 0.4em\relax
  Birkhauser, 2001, pp. 283--324.

\bibitem{Junge_Osinga}
O.~Junge and H.~Osinga, ``A set oriented approach to global optimal control,''
  \emph{ESAIM: Control, Optimisation and Calculus of Variations}, vol.~10,
  no.~2, pp. 259--270, 2004.

\bibitem{Mezic_comparison}
I.~Mezi\'{c} and A.~Banaszuk, ``Comparison of systems with complex behavior,''
  \emph{Physica D}, vol. 197, pp. 101--133, 2004.

\bibitem{Dellnitztransport}
M.~Dellnitz, O.~Junge, W.~S. Koon, F.~Lekien, M.~Lo, J.~E. Marsden, K.~Padberg,
  R.~Preis, S.~D. Ross, and B.~Thiere, ``Transport in dynamical astronomy and
  multibody problems,'' \emph{International Journal of Bifurcation and Chaos},
  vol.~15, pp. 699--727, 2005.

\bibitem{mezic2005spectral}
I.~Mezi{\'c}, ``Spectral properties of dynamical systems, model reduction and
  decompositions,'' \emph{Nonlinear Dynamics}, vol.~41, no. 1-3, pp. 309--325,
  2005.

\bibitem{Mehta_comparsion_cdc}
P.~G. Mehta and U.~Vaidya, ``On stochastic analysis approaches for comparing
  dynamical systems,'' in \emph{Proceeding of IEEE Conference on Decision and
  Control}, Spain, 2005, pp. 8082--8087.

\bibitem{Vaidya_TAC}
U.~Vaidya and P.~G. Mehta, ``Lyapunov measure for almost everywhere
  stability,'' \emph{IEEE Transactions on Automatic Control}, vol.~53, no.~1,
  pp. 307--323, 2008.

\bibitem{raghunathan2014optimal}
A.~Raghunathan and U.~Vaidya, ``Optimal stabilization using lyapunov
  measures,'' \emph{IEEE Transactions on Automatic Control}, vol.~59, no.~5,
  pp. 1316--1321, 2014.

\bibitem{susuki2011nonlinear}
Y.~Susuki and I.~Mezic, ``Nonlinear koopman modes and coherency identification
  of coupled swing dynamics,'' \emph{IEEE Transactions on Power Systems},
  vol.~26, no.~4, pp. 1894--1904, 2011.

\bibitem{mezic_koopmanism}
M.~Budisic, R.~Mohr, and I.~Mezic, ``Applied koopmanism,'' \emph{Chaos},
  vol.~22, pp. 047\,510--32, 2012.

\bibitem{mezic_koopman_stability}
A.~Mauroy and I.~Mezic, ``A spectral operator-theoretic framework for global
  stability,'' in \emph{Proc. of IEEE Conference of Decision and Control},
  Florence, Italy, 2013.

\bibitem{surana_observer}
A.~Surana and A.~Banaszuk, ``Linear observer synthesis for nonlinear
  systemsusing koopman operator framework,'' in \emph{{Proceedings of IFAC
  Symposium on Nonlinear Control Systems}}, Monterey, California, 2016.

\bibitem{yeung2015global}
E.~Yeung, J.~Kim, J.~Gon{\c{c}}alves, and R.~M. Murray, ``Global network
  identification from reconstructed dynamical structure subnetworks:
  Applications to biochemical reaction networks,'' in \emph{Decision and
  Control (CDC), 2015 IEEE 54th Annual Conference on}.\hskip 1em plus 0.5em
  minus 0.4em\relax IEEE, 2015, pp. 881--888.

\bibitem{yeung2018koopman}
E.~Yeung, Z.~Liu, and N.~O. Hodas, ``A koopman operator approach for computing
  and balancing gramians for discrete time nonlinear systems,'' in \emph{2018
  Annual American Control Conference (ACC)}.\hskip 1em plus 0.5em minus
  0.4em\relax IEEE, 2018, pp. 337--344.

\bibitem{yeung2017learning}
E.~Yeung, S.~Kundu, and N.~Hodas, ``Learning deep neural network
  representations for koopman operators of nonlinear dynamical systems,''
  \emph{arXiv preprint arXiv:1708.06850}, 2017.

\bibitem{sparse_Koopman_acc}
S.~Sinha, U.~Vaidya, and E.~Yeung, ``On computation of koopman operator from
  sparse data,'' in \emph{2019 American Control Conference (ACC)}.\hskip 1em
  plus 0.5em minus 0.4em\relax IEEE, 2019, pp. 5519--5524.

\bibitem{johnson2018class}
C.~A. Johnson and E.~Yeung, ``A class of logistic functions for approximating
  state-inclusive koopman operators,'' in \emph{2018 Annual American Control
  Conference (ACC)}.\hskip 1em plus 0.5em minus 0.4em\relax IEEE, 2018, pp.
  4803--4810.

\bibitem{optimal_placement_ECC}
S.~Sinha, U.~Vaidya, and R.~Rajaram, ``Optimal placement of actuators and
  sensors for control of nonequilibrium dynamics,'' in \emph{Control Conference
  (ECC), 2013 European}.\hskip 1em plus 0.5em minus 0.4em\relax IEEE, 2013, pp.
  1083--1088.

\bibitem{optimal_placement_JMAA}
------, ``Operator theoretic framework for optimal placement of sensors and
  actuators for control of nonequilibrium dynamics,'' \emph{Journal of
  Mathematical Analysis and Applications}, vol. 440, no.~2, pp. 750--772, 2016.

\bibitem{dellnitz2002set}
M.~Dellnitz and O.~Junge, ``Set oriented numerical methods for dynamical
  systems,'' \emph{Handbook of dynamical systems}, vol.~2, pp. 221--264, 2002.

\bibitem{DMD_schmitt}
P.~J. Schmid, ``Dynamic mode decomposition of numerical and experimental
  data,'' \emph{Journal of Fluid Mechanics}, vol. 656, pp. 5--28, 2010.

\bibitem{rowley2009spectral}
C.~W. Rowley, I.~Mezi{\'c}, S.~Bagheri, P.~Schlatter, and D.~S. Henningson,
  ``Spectral analysis of nonlinear flows,'' \emph{Journal of fluid mechanics},
  vol. 641, pp. 115--127, 2009.

\bibitem{EDMD_williams}
M.~O. Williams, I.~G. Kevrekidis, and C.~W. Rowley, ``A data--driven
  approximation of the koopman operator: Extending dynamic mode
  decomposition,'' \emph{Journal of Nonlinear Science}, vol.~25, no.~6, pp.
  1307--1346, 2015.

\bibitem{mezic_stochastic_koopman_spectrum}
N.~Crnjaric-Zic, S.~Macesic, and I.~Mezic, ``Koopman operator spectrum for
  random dynamical system,'' \emph{arXiv preprint arXiv:1711.03146}, 2017.

\bibitem{PhysRevE.96.033310}
\BIBentryALTinterwordspacing
N.~Takeishi, Y.~Kawahara, and T.~Yairi, ``Subspace dynamic mode decomposition
  for stochastic koopman analysis,'' \emph{Phys. Rev. E}, vol.~96, p. 033310,
  Sep 2017. [Online]. Available:
  \url{https://link.aps.org/doi/10.1103/PhysRevE.96.033310}
\BIBentrySTDinterwordspacing

\bibitem{robust_DMD_ACC}
S.~Sinha, B.~Huang, and U.~Vaidya, ``Robust approximation of koopman operator
  and prediction in random dynamical systems,'' in \emph{2018 Annual American
  Control Conference (ACC)}.\hskip 1em plus 0.5em minus 0.4em\relax IEEE, 2018,
  pp. 5491--5496.

\bibitem{robust_DMD_arxiv}
S.~Sinha, H.~Bowen, and U.~Vaidya, ``On robust computation of koopman operator
  and prediction in random dynamical systems,'' \emph{arXiv preprint
  arXiv:1803.08562}, 2018.

\bibitem{Lasota}
A.~Lasota and M.~C. Mackey, \emph{Chaos, Fractals, and Noise: Stochastic
  Aspects of Dynamics}.\hskip 1em plus 0.5em minus 0.4em\relax New York:
  Springer-Verlag, 1994.

\bibitem{korda_mezic_predictor}
M.~Korda and I.~Mezi{\'c}, ``Linear predictors for nonlinear dynamical systems:
  Koopman operator meets model predictive control,'' \emph{arXiv preprint
  arXiv:1611.03537}, 2016.

\bibitem{KUTLUAY1999251}
S.~Kutluay, A.~Bahadir, and A.~Ozdeş, ``Numerical solution of one-dimensional
  burgers equation: explicit and exact-explicit finite difference methods,''
  \emph{Journal of Computational and Applied Mathematics}, vol. 103, no.~2, pp.
  251 -- 261, 1999.

\end{thebibliography}

\end{document}